\documentclass{emulateapj}
\usepackage{psfig,natbib,amsmath}

\newcommand{\citepeg}[1]{\citep[{e.g.,}][]{#1}}
\def\lsim{\hbox{ \rlap{\raise 0.425ex\hbox{$<$}}\lower 0.65ex\hbox{$\sim$}}}
\def\gsim{\hbox{ \rlap{\raise 0.425ex\hbox{$>$}}\lower 0.65ex\hbox{$\sim$}}}

\def\arcsec{\hbox{$^{\prime\prime}$}}
\def\arcdeg{\mbox{$^\circ$}}%
\newcommand{\ubv}{\protect\hbox{$U\!BV$} }
\newcommand{\ubvri}{\protect\hbox{$U\!BV\!RI$} }
\newcommand{\vri}{\protect\hbox{$V\!RI$} }

\begin{document}

\title{The Troublesome Broadband Evolution of GRB 061126: 
       Does a Grey Burst Imply Grey Dust?}

\def\berk{1}
\def\lick{2}
\def\nmsu{3}
\def\harvard{4}
\def\lanl{5}
\def\ssl{6}
\def\chicago{7}

\author{D.~A.~Perley\altaffilmark{\berk},
		J.~S.~Bloom\altaffilmark{\berk},		
		N.~R.~Butler\altaffilmark{\berk},
		L.~K.~Pollack\altaffilmark{\lick},
        J.~Holtzman\altaffilmark{\nmsu},
		C.~H.~Blake\altaffilmark{\harvard},
		D.~Kocevski\altaffilmark{\berk},
		W.~T.~Vestrand\altaffilmark{\lanl},
		W.~Li\altaffilmark{\berk},
		R.~J.~Foley\altaffilmark{\berk},
		E.~Bellm\altaffilmark{\ssl},
 		H.-W.~Chen\altaffilmark{\chicago},
		J.~X.~Prochaska\altaffilmark{\lick},
		D.~Starr\altaffilmark{\berk},
		A.~V.~Filippenko\altaffilmark{\berk},
        E.~E.~Falco\altaffilmark{\harvard},
        A.~H.~Szentgyorgyi\altaffilmark{\harvard},
		J.~Wren\altaffilmark{\lanl}, 
		P.~R.~Wozniak\altaffilmark{\lanl}, 
		R.~White\altaffilmark{\lanl}, and
		J.~Pergande\altaffilmark{\lanl}
}

\altaffiltext{\berk}{Department of Astronomy, 
        University of California, Berkeley, CA 94720-3411.}
\altaffiltext{\lick}{University of California Observatories/Lick Observatory,
         University of California, Santa Cruz, CA 95064.}
\altaffiltext{\nmsu}{Department of Astronomy, New Mexico State University
         P. O. Box 30001, MSC 4500,  Las Cruces, New Mexico 88003-8001.}
\altaffiltext{\harvard}{Harvard-Smithsonian Center for Astrophysics, 60 Garden Street,
          Cambridge, MA 02138.}
\altaffiltext{\lanl}{Los Alamos National Laboratory,
         Los Alamos, NM, 87545.}
\altaffiltext{\ssl}{University of California, Berkeley, Space Sciences Laboratory, 
        7 Gauss Way, Berkeley CA 94720.}
\altaffiltext{\chicago}{Department of Astronomy and Astrophysics, University of Chicago, 
          5640 S. Ellis Ave, Chicago, IL 60637.}


\begin{abstract}
We report on observations of a gamma-ray burst (GRB\,061126) with an extremely bright ($R \approx 12$ mag at peak) early-time optical afterglow.  The optical afterglow is already fading as a power law 22 seconds after the trigger, with no detectable prompt contribution in our first exposure, which was coincident with a large prompt-emission gamma-ray pulse.  The optical--infrared photometric spectral energy distribution is an excellent fit to a power law, but it exhibits a moderate red-to-blue evolution in the spectral index at about 500 s after the burst. This color change is contemporaneous with a switch from a relatively fast decay to slower decay.  The rapidly decaying early afterglow is broadly consistent with synchrotron emission from a reverse shock, but a bright forward-shock component predicted by the intermediate- to late-time X-ray observations under the assumptions of standard afterglow models is not observed. Indeed, despite its remarkable early-time brightness, this burst would qualify as a dark burst at later times on the basis of its nearly flat optical-to-X-ray spectral index.  Our photometric spectral energy distribution provides no evidence of host-galaxy extinction, requiring either large quantities of grey dust in the host system (at redshift $1.1588 \pm 0.0006$, based upon our late-time Keck spectroscopy) or separate physical origins for the X-ray and optical afterglows. 
\end{abstract}


\section{Introduction}

While the study of the early-time X-ray afterglows of gamma-ray bursts (GRBs) has seen enormous strides since the launch of the \emph{Swift} satellite \citep{Gehrels+2004}, progress in the understanding of the longer-wavelength emission has been somewhat more measured.  This is unfortunate, as a complete understanding of \emph{Swift} afterglows can only come from a combined broadband picture that allows us to systematically investigate whether the peculiarities seen in X-ray data carry over into the optical domain.  Many of the same questions raised by recent X-ray results can also be asked about the optical.  Is there a prompt optical component, analogous to the steeply decaying component seen in X-rays \citep{Barthelmy+2005}?  Does the optical light curve show unusual features suggestive of energy injection, such as the nearly ubiquitous X-ray shallow-decay phase \citep{Nousek+2006}?  Are there achromatic optical and X-ray breaks?  Do the optical and X-ray afterglows even have a common origin at early times?

Previous studies have provided important hints.  Most observations have been interpreted to support the consensus picture of synchrotron emission originating from a forward shock as it sweeps through the interstellar medium (e.g., \citealt{DaiLu1999}; \citealt{Vrba+2000}), or less commonly through a stellar wind \citep{Price+2002, Nysewander+2006} --- see \citet{Chevalier2007} for a review.  In a smaller number of cases \citep{Akerlof+1999, Li+2003b, KobayashiZhang2003, ShaoDai2005}, very early-time data have provided tentative evidence for an additional emission component originating from the reverse shock as it travels backward into the shocked material in the frame of the forward shock.  Recently, studies of early-time light curves have also shown evidence of significant delay between the onset of the prompt emission and the afterglow \citep{Rykoff+2004}, and in at least one case complicated energy injection activity as late as nearly an hour after the gamma-ray burst \citep{Wozniak+2006}, long after the gamma-ray emission has faded away.  However, simultaneous, correlated optical and gamma-ray emission has also been reported \citep{Blake+2005, Vestrand+2005, Vestrand+2006, Yost+2007} for some events.  In the \emph{Swift} era, comparison to the very early X-ray afterglow has also been of great interest (e.g., \citealt{Quimby+2006}).

Most interpretations of the early afterglow have been based on unfiltered observations, or observations in a single filter.  Without information about the frequency domain, the reported early-time behaviors discussed above are difficult to definitively associate with any single physical interpretation.  Fortunately, the increasing number of fast-responding robotic ground-based observatories, the maturation of existing ones, and the rapid-response capabilities of \emph{Swift} are beginning to address this observational gap.

In the following discussion, we report on one of the brightest bursts of the \emph{Swift} era, GRB\,061126.  The breadth and rapidity of the ground-based response to this burst were remarkable, including unfiltered detection during the prompt emission and multi-color simultaneous detections in filters from $U$ through $K_s$ (ranging a full decade in frequency) starting less than one minute after the burst trigger.  This data set provides the opportunity to examine in unprecedented detail the time-dependent color properties of an early GRB afterglow.   

In \S \ref{sec:obs} we present our observations from infrared (IR) through gamma rays of the early afterglow and our late-time Keck spectrum of the host, establishing the probable redshift of this system to be $z=1.1588$.  In \S \ref{sec:prompt} we examine the properties of the prompt emission, and show that the high-energy and optical emission are observationally uncorrelated temporally or spectrally, even at very early times.  In \S \ref{sec:optlc}--\ref{sec:optcolor} we examine the properties of the optical--IR light curve, and provide evidence for a red-to-blue change in the spectral index of $\Delta\beta \approx 0.3$ at early times.  We investigate the X-ray behavior in \S \ref{sec:closure}, and show that no standard adiabatic model can fully explain the behavior seen by the X-Ray Telescope (XRT).  Finally, while in \S \ref{sec:revshock}--\ref{sec:optlate} we show that the earliest afterglow appears reasonably fit by a reverse shock and the later afterglow by a forward shock based on the optical data alone, in \S \ref{sec:bbfits} we demonstrate that an extrapolation of the X-ray spectrum overpredicts the contemporaneous optical flux by a factor of 5--20. We demonstrate using the optical--IR spectral energy distribution (SED) that this discrepancy cannot be due to any known dust extinction law.  Unless we appeal to large quantities of grey dust, a possibility we discuss in \S \ref{sec:greydust}, we argue in \S \ref{sec:compton}--\ref{sec:separate} that the X-ray and optical afterglow emission from this burst have separate physical origins.

\section{Observations}
\label{sec:obs}

\subsection{Swift BAT and XRT}

At 08:47:56 on 26 November 2006 (UT dates are used throughout this paper), the \emph{Swift} Burst Alert Telescope (BAT) triggered and located GRB\,061126.  Unfortunately, due to an Earth-limb constraint, \emph{Swift} was unable to slew promptly to the target for 23 minutes and could not begin observations with the XRT or the Ultraviolet/Optical Telescope (UVOT) before that time. After 23 minutes, \emph{Swift} slewed to the burst position and detected a fading X-ray afterglow \citep{GCN5855}. 
 
We download the \emph{Swift} BAT and XRT data from the {\it Swift}~Archive\footnote{ftp://legacy.gsfc.nasa.gov/swift/data .}. The XRT data are processed with version 0.10.3 of the {\tt xrtpipeline} reduction script from the HEAsoft~6.0.6\footnote{http://heasarc.gsfc.nasa.gov/docs/software/lheasoft/ .} software release.  We employ the latest (19 December 2006) XRT and BAT calibration files.  We establish the BAT energy scale and mask weighting by running the {\tt bateconvert} and {\tt batmaskwtevt} tasks, also from the HEAsoft~6.0.6 software release.  BAT spectra and light curves are extracted with the {\tt batbinevt} task, and response matrices are produced by running {\tt batdrmgen}.  We apply the systematic error corrections to the low-energy BAT spectral data as advised by the BAT Digest website\footnote{http://swift.gsfc.nasa.gov/docs/swift/analysis/bat\_digest.html .}. The spectral normalizations are corrected for satellite slews using the 
{\tt batupdatephakw} task. 

The reduction of XRT data from cleaned event lists output by {\tt xrtpipeline} to science-ready light curves and spectra is described in detail in \cite{ButlerKocevski2007}.  The XRT, BAT, and RHESSI (Reuven Ramaty High Energy Solar Spectroscopic Imager) data are fit using ISIS\footnote{http://space.mit.edu/CXC/ISIS/ .}. The XRT light curve is converted to normalized 100 keV flux using a scaling of 7.5 $\mu$Jy/cps.


\subsection{RHESSI}

RHESSI \citep{Lin+2002} is a dedicated solar observatory which uses nine germanium detectors to image the Sun at hard X-ray to gamma-ray energies (3 keV -- 17 MeV).  These detectors are unshielded and therefore frequently detect emission from off-axis GRBs.  GRB\,061126 was detected by RHESSI, which with its large spectral range allows us to complete the high-energy spectrum of this event.

To model the RHESSI response to off-axis photons, we have used Monte Carlo simulations and a detailed mass model.  Since RHESSI rotates about its axis with a 4~s period, we have generated azimuthally averaged responses spaced $15^\circ$  apart in polar angle.  These responses are two-dimensional matrices of effective area: input photon energy vs.\ detected count energy bins.  At present both energy axes are 64 logarithmic bins from 10 keV to 10 MeV.  We generate the response matrices with {\tt MGEANT} \citep{Sturner+2000} simulations: each response requires 64 simulations of a monoenergetic input spectrum, one for each photon energy bin.  For an individual GRB, we generate and subtract a background count spectrum using data intervals before and after the burst.  We generate a burst-specific response matrix with a weighted average of the two adjacent $15^\circ$ responses.  Convolving a spectral model with the response yields a model count spectrum for fitting to the GRB data.

\subsection{RAPTOR}

\begin{deluxetable}{lccll}
\tabletypesize{\small}
\tablecaption{RAPTOR Observations of GRB\,061126\label{tab:raptorphot}}
\tablecolumns{6}
\tablehead{
\colhead{$\Delta t$\tablenotemark{a}} & \colhead{Filter} &
\colhead{Exp.~Time} &
\colhead{Magnitude\tablenotemark{b}} & \colhead{Flux\tablenotemark{b}} \\
\colhead{(s)} & \colhead{} &
\colhead{(s)} & \colhead{} &
\colhead{($\mu$Jy)}}
\startdata
   23.4 & {\rm clear} &     5.0 & $ 12.26 \pm  0.01\tablenotemark{c}$ & $38725.8 \pm   355.0$ \\
   32.3 & {\rm clear} &     5.0 & $ 12.66 \pm  0.02$ & $26791.7 \pm   489.0$ \\
   41.1 & {\rm clear} &     5.0 & $ 13.08 \pm  0.03$ & $18197.0 \pm   495.9$ \\
\enddata 
\tablenotetext{a}{Exposure mid-time, measured from the \emph{Swift} trigger (UT 08:47:56.4).}
\tablenotetext{b}{Observed value; not corrected for Galactic extinction.}
\tablenotetext{c}{Contemporaneous with strong GRB pulse; point not used in modeling.}
\tablecomments{Only the first three observations are given here.  Additional data are available in the online supplement, or by request.}
\end{deluxetable}

The RAPTOR (Rapid Telescopes for Optical Response) experiment \citep{Vestrand+2002} consists of a series of small telescopes used to conduct transient surveys as well as perform rapid followup of GRBs and other events.  One of these telescopes, RAPTOR-S, is a 0.4~m fully autonomous robotic telescope, typically operated at focal ratio $f$/5. It is equipped with a 1k $\times$ 1k pixel CCD camera employing a back-illuminated Marconi CCD47-10 chip with 13 $\mu$m pixels. The telescope is owned by Los Alamos National Laboratory and located at the Fenton Hill Observatory (106.67$\arcdeg$ W, 35.88$\arcdeg$ N) at an altitude of $\sim$2500 m in the Jemez Mountains of New Mexico.

RAPTOR-S responded to the \emph{Swift} trigger at 08:48:17.29, or 20.87~s after the trigger and 4.3~s after receiving the GCN (Gamma-ray Burst Coordination Network) packet.  The telescope took a series of nine unfiltered 5~s exposures (the first two of which occurred while detectable gamma-ray emission was still ongoing), followed by a series of 10~s and 30~s exposures.  The optical transient is detected in all these frames. Preliminary photometric calibration was performed using the $R$-band magnitudes from the USNO (United States Naval Observatory) B1.0 catalog.  However, for consistency with the unfiltered KAIT (Katzman Automatic Imaging Telescope) observations, which were calibrated using the more precise SDSS (Sloan Digital Sky Survey) measurements, we subtract a constant offset of 0.16 mag post-calibration.  The RAPTOR photometry (not including this final offset) is given in Table \ref{tab:raptorphot}.

\subsection{PAIRITEL}

\begin{figure*}[t] 
\centerline{\includegraphics[width=5.5in,angle=0]{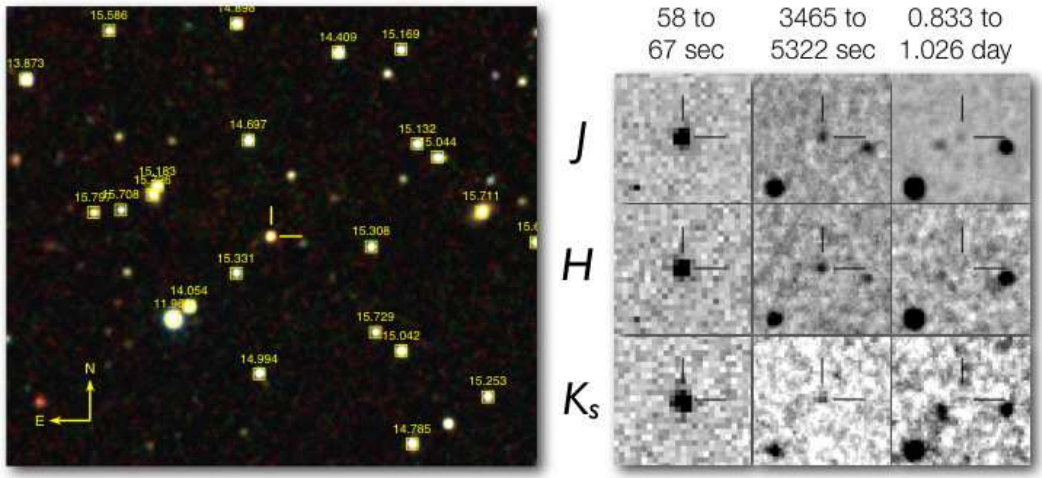}}
\caption[relation] {\small False-color PAIRITEL finding chart (260\arcsec\ $\times$ 260\arcsec) of the afterglow of GRB\,061226 (left). The 2MASS catalog stars used for the photometric calibration are denoted with boxes at the catalog positions and labeled with $J$-band (2MASS) magnitudes. At right, IR images show the fading of the afterglow from 58~s to 1~d after the GRB; the images are progressively deeper at later epochs. }
\label{fig:ptel}
\end{figure*}

\begin{deluxetable}{lccll}
\tabletypesize{\small}
\tablecaption{PAIRITEL Observations of GRB\,061126\label{tab:ptelphot}}
\tablecolumns{6}
\tablehead{
\colhead{$\Delta t$\tablenotemark{a}} & \colhead{Filter} &
\colhead{Exp.~Time} &
\colhead{Magnitude\tablenotemark{b}} & \colhead{Flux\tablenotemark{b}} \\
\colhead{(s)} & \colhead{} &
\colhead{(s)} & \colhead{} &
\colhead{($\mu$Jy)}}
\startdata
   64.8 & $J$ &     4.3 & $ 11.72 \pm  0.02$ & $32658.7 \pm   507.4$ \\
  262.0 & $H$ &     4.0 & $ 12.89 \pm  0.05$ & $ 7177.9 \pm   304.1$ \\
  136.7 & $K_S$ &     4.5 & $ 11.01 \pm  0.05$ & $26278.5 \pm  1298.0$ \\
\enddata 
\tablecomments{Only the first usable exposure in each band is given.  The full table of photometry is available upon request, or in our online supplement.}
\tablenotetext{a}{Exposure mid-time, measured from the \emph{Swift} trigger (UT 08:47:56.4).}
\tablenotetext{b}{Observed value; not corrected for Galactic extinction.}
\end{deluxetable}

Starting in 2003, we began to automate the 1.3-m Peters Telescope, formerly used for the Two Micron All Sky Survey \citep[2MASS;][]{Skrutskie+2006}, on Mt.\ Hopkins, Arizona.  The telescope was re-outfitted with the southern 2MASS camera, and all of the control and data acquisition systems were rewritten \citep{Bloom+2006c}. The Peters Automated Infrared Imaging Telescope (PAIRITEL) has been obtaining useful IR observations of GRBs since 2004.

At 08:48:18 ($t = 22$ s), PAIRITEL was triggered with the GCN BAT position notice of GRB\,061126.  The autonomous slew of the telescope and dome began at 08:48:22 and ended at 08:48:47; the slew time was short since we had been observing M82 (23.9\arcdeg\ to the east of the GRB) immediately prior to the GRB.  After an initial reset of the camera, the first 7.8~s images in $J$, $H$, and $K_s$ bands were obtained starting at 08:48:54.35 ($t = 58$ s). We continued with a dense sampling of observations over the next three hours as well as several hours of imaging the following night.

Reductions of the individual images were performed using a set of customized scripts written in {\sc pyraf} and Python. The afterglow was well detected with signal-to-noise ratio (S/N) $>10$ in individual images for the first 20 min of observations (Figure \ref{fig:ptel}). In fact, the $H$- and $K_s$-band fluxes of the afterglow are so bright in the first few minutes that the pixel responses were in the nonlinear regime.   Unfortunately, the cloud cover in Arizona was highly variable during the first 30 min of GRB\,061126 observations, leading to variable transmission on 10~s timescales.\footnote{An animated image showing the variable transmission and fading afterglow can be viewed at http://lyra.berkeley.edu/\~{}jbloom/grb061126a-hband.mpg .} As such, in our analysis we refit the zeropoint in every individual exposure to the 2MASS catalog. The typical root-mean-square (rms) uncertainties in the zeropoints are 2--3\%. Given the large variations in the sub-pixel response function for NICMOS3 arrays, we have found that aperture photometry on individual exposures suffers a roughly 3\% systematic uncertainty from image to image (Blake et al.\ 2007, in preparation). Table \ref{tab:ptelphot} gives the aperture magnitude measurements from the PAIRITEL observations.  In this table and in all plots and modeling, we exclude exposures in which the CCD response was nonlinear, as well as $H$-band observations during periods of poor transmission.

We determined the position of the afterglow to be ($\alpha, \delta$) = (05$^h$46$^m$24.47$^s$ $\pm$ 0.16\arcsec, +64$^\circ$12$'$38.60\arcsec $\pm$ 0.18\arcsec) (J2000) from a stacked $J$-band image covering the first 30 min after the trigger. The quoted uncertainties are 1$\sigma$, dominated by the astrometric mapping error from our stacked image to the catalog positions of 90 2MASS sources.

\subsection{NMSU 1 m Telescope}

\begin{deluxetable}{lccll}
\tabletypesize{\small}
\tablecaption{NMSU 1~m Observations of GRB\,061126\label{tab:nmsuphot}}
\tablecolumns{6}
\tablehead{
\colhead{$\Delta t$\tablenotemark{a}} & \colhead{Filter} &
\colhead{Exp.~Time} &
\colhead{Magnitude\tablenotemark{b}} & \colhead{Flux\tablenotemark{b}} \\
\colhead{(s)} & \colhead{} &
\colhead{(s)} & \colhead{} &
\colhead{($\mu$Jy)}}
\startdata
   52.3 & $I$ &    10.0 & $ 12.53 \pm  0.00$ & $23746.5 \pm    65.5$ \\
   96.8 & $R$ &    10.0 & $ 14.22 \pm  0.01$ & $ 6356.2 \pm    35.0$ \\
  148.8 & $V$ &    20.0 & $ 15.32 \pm  0.01$ & $ 2737.8 \pm    17.6$ \\
  213.2 & $B$ &    40.0 & $ 16.43 \pm  0.01$ & $ 1108.7 \pm     9.2$ \\
  302.9 & $U$ &    60.0 & $ 16.67 \pm  0.03$ & $  411.9 \pm     9.4$ \\
\enddata 
\tablecomments{Only the first filter cycle is given.  Additional data are available in the
 online supplement, or by request.}
\tablenotetext{a}{Exposure mid-time, measured from the \emph{Swift} trigger (UT 08:47:56.4).}
\tablenotetext{b}{Observed value; not corrected for Galactic extinction.}
\end{deluxetable}

Optical observations in the Johnson-Cousins \ubvri filters were obtained using the New Mexico State University robotic 1~m telescope located at Apache Point Observatory.  Because the telescope happened to be pointed relatively near the burst location, the first observations were started only 47~s after the burst trigger, and only 31~s after the alert.  The telescope took five sequences of observations in the order $I,R,V,B,U$ with exposure times of 10, 10, 20, 40, and 60~s, respectively, and then took another five sequences of observations with exposure times of 20, 20, 40, 80, and 120~s. With overhead, these ten sequences took about an hour to complete. The afterglow was detected in all of the observations, with random errors ranging from better than 0.01 mag in the first series to 0.1--0.2 mag in the last series.

The afterglow brightness was measured using aperture photometry with an aperture of 3\arcsec\ radius; several reference stars in the field were also measured. Calibration was achieved via observations of these reference stars, along with \ubvri standard stars, on 22 December 2006. A standard photometric solution was derived for this night, yielding calibrated magnitudes for the reference stars. On the calibration night, the standard-star solution yielded rms deviations of about 0.03 mag in each bandpass, so the calibration zeropoints are accurate only to this level.
The final photometry is given in Table \ref{tab:nmsuphot}.

\subsection{KAIT and the Lick Nickel Telescope}

\begin{deluxetable}{lccll}
\tabletypesize{\small}
\tablecaption{KAIT and Lick 1~m Observations of GRB\,061126\label{tab:lickphot}}
\tablecolumns{6}
\tablehead{
\colhead{$\Delta t$\tablenotemark{a}} & \colhead{Filter} &
\colhead{Exp.~Time} &
\colhead{Magnitude\tablenotemark{b}} & \colhead{Flux\tablenotemark{b}} \\
\colhead{(s)} & \colhead{} &
\colhead{(s)} & \colhead{} &
\colhead{($\mu$Jy)}}
\startdata
  366.0 & clear &20.0 & $ 16.29 \pm  0.12$ & $  946.2 \pm    99.0$ \\
  409.5 &  $V$ &    45.0 & $ 17.25 \pm  0.12$ & $  464.5 \pm    48.6$ \\
  465.5 &  $I$ &    45.0 & $ 16.24 \pm  0.11$ & $  776.2 \pm    74.8$ \\
 3336.0 &  $R$ &   300.0 & $ 18.86 \pm  0.05$ & $   88.7 \pm     4.0$ \\
\enddata 
\tablecomments{$R$-band points are from the Lick 1~m (Nickel) telescope; all others are from KAIT.
 Only the first exposure in each filter is given; additional data are available in the online
 supplement, or by request.}
\tablenotetext{a}{Exposure mid-time, measured from the \emph{Swift} trigger (UT 08:47:56.4).}
\tablenotetext{b}{Observed value; not corrected for Galactic extinction.}
\end{deluxetable}

\begin{deluxetable}{lccll}
\tabletypesize{\small}
\tablecaption{UVOT Observations of GRB\,061126\label{tab:uvotphot}}
\tablecolumns{6}
\tablehead{
\colhead{$\Delta t$\tablenotemark{a}} & \colhead{Filter} &
\colhead{Exp.~Time} &
\colhead{Magnitude\tablenotemark{b}} & \colhead{Flux\tablenotemark{b}} \\
\colhead{(s)} & \colhead{} &
\colhead{(s)} & \colhead{} &
\colhead{($\mu$Jy)}}
\startdata
 2152.0 & $U$ &   967.5 & $ 18.64 \pm  0.11$ & $   67.1 \pm     6.6$ \\
 2114.0 & $B$ &   809.5 & $ 19.05 \pm  0.12$ & $   99.2 \pm    10.1$ \\
 2178.0 & $V$ &   809.5 & $ 18.43 \pm  0.14$ & $  156.5 \pm    19.1$ \\
 8862.0 & $V$ &   902.0 & $ 20.51 \pm  0.32$ & $   23.0 \pm     5.8$ \\
 7700.3 & $U$ &   196.6 & $ 19.75 \pm  0.19$ & $   24.2 \pm     3.8$ \\
15119.5 & $U$ &   295.1 & $ 20.33 \pm  0.38$ & $   14.2 \pm     4.2$ \\
 2807.3 & $B$ &   196.6 & $ 19.52 \pm  0.12$ & $   64.4 \pm     7.0$ \\
 7905.3 & $B$ &   196.6 & $ 20.61 \pm  0.26$ & $   23.5 \pm     5.0$ \\
 3421.3 & $V$ &   196.6 & $ 19.07 \pm  0.23$ & $   87.2 \pm    16.5$ \\
\enddata
 \tablecomments{UVOT data points were not used in our light-curve models. $U$, $B$, and $V$ filter
 measurements are our own re-reductions.  UV and white filter photometry are from GCN Report 16.2
 \citep{GCNR16.2}; they are not repeated here, but are also available in our online supplement.}
\tablenotetext{a}{Exposure mid-time, measured from the \emph{Swift} trigger (UT 08:47:56.4).}
\tablenotetext{b}{Observed value; not corrected for Galactic extinction.}
\end{deluxetable}

The 0.76~m robotic Katzman Automatic Imaging Telescope (KAIT; \citealt{Filippenko+2001}) and its GRB alert system \citep{Li+2003a} responded to the GCN notice with the position of GRB\,061126 at $t=16$~s after the trigger, and attempted to execute a pre-arranged observation sequence. Unfortunately, the weather conditions were poor, so KAIT did not acquire a useful image until $t = 305$~s. Also, the telescope did not have a successful focusing procedure before the GRB observations (again due to bad weather), so the images were not fully in focus. Nevertheless, a sequence of $V$, $I$, and unfiltered observations was made, and the GRB afterglow was detected in most images. A successful focusing procedure was executed during the middle of the GRB observations, and KAIT followed the GRB until humidity forced the system to shut down at $t\approx$ 1.8~hr. The images were automatically processed with the proper dark current, bias, and flat fields before measuring photometry and calibrating relative to ten SDSS stars \citep{GCN5985, DR4}.  The reference magnitudes of these stars were converted to \vri using the transformation equations of Lupton (2005)\footnote{http://www.sdss.org/dr4/algorithms/sdssUBVRITransform.html\#Lupton2005 .}.  Unfiltered observations were calibrated using the $R$-band magnitudes.

A sequence of 300~s $R$-band images was also manually obtained at the Lick Observatory Nickel 1~m telescope from $t\approx$ 1.0~hr to 1.7~hr.  The observations were again terminated prematurely due to the weather conditions. The images were manually reduced with the proper calibration files (bias, dark current, and flat-field images), and calibrated relative to our SDSS reference stars.  The photometry is given in Table \ref{tab:lickphot}.

\subsection{{\it Swift} UVOT}

\emph{Swift} began follow-up observations of GRB\,061126 at 1605~s after the  burst trigger, its slew having been delayed as a result of the Earth-limb constraint.  Despite this time delay the afterglow was still detected, albeit marginally, in all of the UVOT filters except for UVW2.  

We acquired the UVOT imaging data from the NASA archive\footnote{http://heasarc.gsfc.nasa.gov/docs/swift/archive/ .}.  Unfortunately the afterglow had already become quite faint by the beginning of the observations, and is not well detected except in stacked exposures.  To calculate the most accurate photometry possible, we therefore bin observations obtained at similar epochs, and perform aperture photometry using the optimal aperture size as given in the prescription of \cite{Li+2006} for the $V$, $B$, and $U$ data.  For the UV filters, we use the photometry reported by the UVOT team in the most recent GCN Report for this burst \citep{GCNR16.2}.  Our photometry is given in Table \ref{tab:uvotphot}.

Despite our small-aperture analysis, the \ubv photometry has large uncertainties, and some points are not formally consistent with effectively simultaneous ground-based  observations in the same filters. This is not necessarily surprising; due to the extremely faint nature of the afterglow by the time that \emph{Swift} completed its slew, the actual uncertainty on these measurements may be significantly larger than the nominal photometric error.  We do not use these points in our modeling, but we do include them in our plots.  The UVW1 and UVM2 points have even larger uncertainties (0.4--0.8 mag), and no ground-based calibration is available; moreover, Galactic extinction (which is significant in this direction: $A_{UVM2} \approx$ 1.8 mag) is increasingly uncertain toward these wavelengths.  Therefore, we exclude these points from the formal fits as well, and restrict our modeling to the much more precise ground-based photometry.  However, we do include the UV points in our SED plots for comparison.

\subsection{GCN Circulars}

For comparative purposes, our plots also include points from the GCN Circulars\footnote{http://gcn.gsfc.nasa.gov/gcn3\_archive.html .}, but we do not actually use these points in our fitting.  Most early-time data were calibrated against the USNO B1.0 survey \citep{Monet+2003}, which does not contain very accurate photometry.  Some data points were also calibrated against a preliminary release of an SDSS pre-burst observation of this field \citep{GCN5863} which was later found to be incorrectly calibrated (indeed, our own use of these observations for  our preliminary calibrations exposed the problem and motivated the re-release of the SDSS calibration used for the KAIT reductions, \citealt{GCN5985}.) The photometry is summarized in Table \ref{tab:gcnphot}.

\begin{figure*}[t] 
\centerline{\includegraphics[width=4.0in,angle=0]{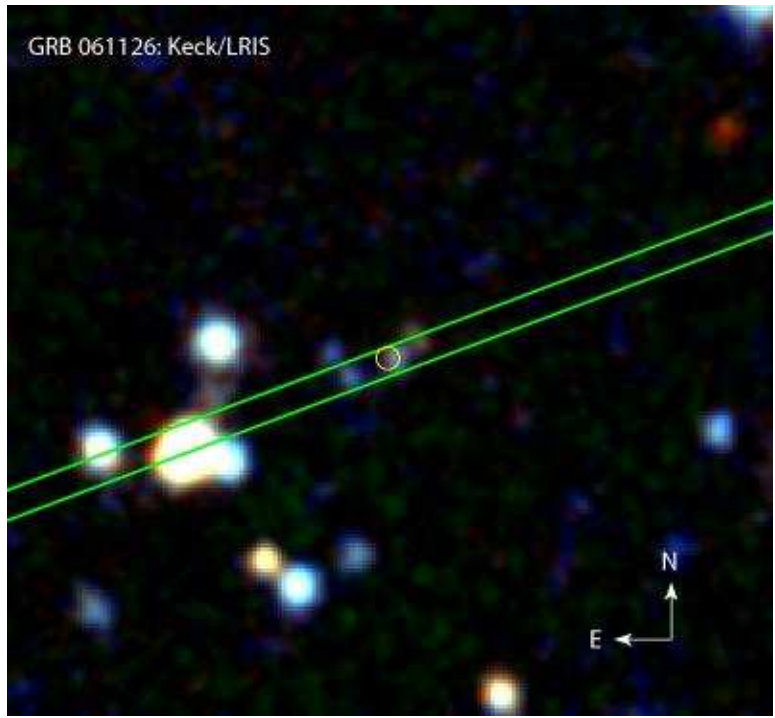}}
\caption[relation] {\small Keck~I (LRIS) late-time finding chart.  The image is 25\arcsec\ on a side and represents the stacked $g'$ and $R$ bands. The placement of the 1\arcsec-wide slit is shown in green. The white circle is the $2\sigma$ position of the afterglow as measured from PAIRITEL images.  The probable host galaxy is visible at this location.  The object is blue and extended, appearing to have a complicated morphology.  Further down the slit 21.1\arcsec\ to the southeast is a relatively bright galaxy with $z=0.6225\pm0.0004$, based on emission from [O~III] $\lambda\lambda$4960.2, 5008.2, $H\beta$, and [O~II] $\lambda\lambda$3727.11, 3729.86.  (This second galaxy is not associated with the GRB.)}
\label{fig:keckim}
\end{figure*}  

\subsection{Keck Host Imaging and Spectroscopy}

The galaxy hosting GRB\,061126 was observed on 18 January 2006 with the Low Resolution Imaging Spectrometer (LRIS; \citealt{Oke+1995}) on the Keck~I 10~m telescope.   A total of 3 $\times$ 300~s of imaging was obtained simultaneously in $R_c$ and $g'$ filters, followed by two long-slit spectroscopic exposures of 20~min each over the host location identified in our images (Figure \ref{fig:keckim}).
  
Imaging data were reduced using standard techniques.  In the final, reduced images we identify an extended source consistent with the GRB afterglow position, apparent in both our $g'$ and $R$ imaging.  This source, which we identify as the GRB host, has significant substructure, with two clearly distinct elongated knots, each of which may have further substructure.  The afterglow is offset by $<1$\arcsec\ from the brightest knot.   The magnitude of the whole complex, using a large (2.55\arcsec\ radius) aperture fixed at the position ($\alpha, \delta$) = 05$^h$46$^m$24.53$^S$, +64$^\circ$12$'$39.31\arcsec) and performing photometric calibration using the subset of our SDSS stars that are unsaturated in the Keck images, is $g = 25.13 \pm 0.16$ and $R = 24.10 \pm 0.11$ mag. After correcting for Galactic extinction, the $g-R$ color is $0.88 \pm 0.19$ mag.

Using the long slit of width 1.0\arcsec, the 600 line mm$^{-1}$ grating blazed at 7500~\AA, and the 300 line mm$^{-1}$ grism blazed at 5000~\AA, we obtained  spectra covering the wavelength range 6680--9266~\AA\ and 3200--7649~\AA\ with the red and blue cameras, respectively. The spectroscopic data were processed with an IDL package\footnote{Specifically, the Longslit codes now bundled within XIDL, http://www.ucolick.org/\~{}xavier/IDL/index.html .} customized for LRIS long-slit reductions developed by J. Hennawi and J. X. Prochaska.  The observed PA of -69.35$^\circ$ is significantly different from the parallactic angle, but as we are only seeking a line identification the effects of differential slit losses \citep{Filippenko1982} are not very significant.

\begin{figure*}[p] 
\centerline{\includegraphics[width=4.0in,angle=0]{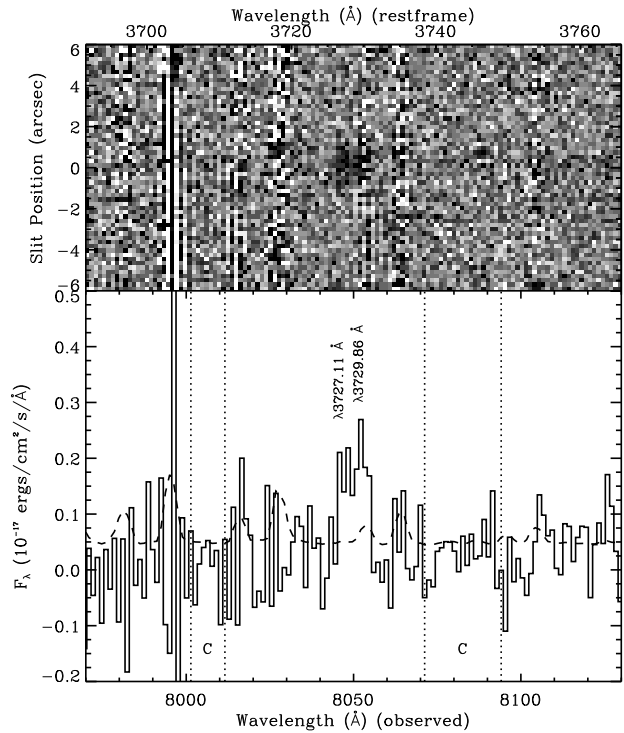}}
\caption[relation] {\small Spectrum of the host galaxy of GRB 061126 in the vicinity of the sole emission line, interpreted as the [\ion{O}{2}] doublet. The data were taken with LRIS on Keck I using the 1\arcsec\ slit and the 600/7500 grism. Top: Background subtracted, two-dimensional spectrogram showing the weighted mean of two 1200~s exposures. Pixels contaminated by cosmic rays in one exposure are excluded from the mean. Bottom: One-dimensional, coadded spectrum. The profile fit from a bright, nearby source was used to extract a spectrum at the known location of the GRB host galaxy. The dashed line represents the $1\sigma$ uncertainty at each pixel. The width of the emission feature is comparable to the spacing of the [\ion{O}{2}] doublet at a redshift of 1.16. The doublet spacing is approximately equal to the instrument's resolution element at 8050~\AA, so we expect the doublet to be barely resolved. We measure the galaxy's continuum by computing the median flux in two regions free of night-sky lines (regions ``C''). Using this continuum value, and measuring the signal between 8042.0 and 8056.0 \AA, we find that the total flux in this emission feature is $(1.6 \pm 0.2) \times 10^{-17}$ erg s$^{-1}$ cm$^{-2}$.}
\label{fig:lrisspec}
\end{figure*}

The two-dimensional reduced spectra show a faint blue continuum and a sole emission feature at a vaccuum wavelength of $\sim8050\,$\AA\ (Figure~\ref{fig:lrisspec}).  From the one-dimensional spectrum it is clear that the emission feature is slightly resolved; it has a full width at half-maximum intensity (FWHM) of roughly 10~\AA, while the spectral resolution in that wavelength range, as determined by the FWHM of arc-line profiles, is $5.8$~\AA.  As we detect only one emission feature, it is impossible to definitively report the redshift of the GRB host.  However, the width of the line and the presence of obvious continuum blueward of this feature suggest that it likely corresponds to the [\ion{O}{2}] $\lambda\lambda$ 3727.11, 3729.86~\AA\ doublet.  Using our best-fit observed line centroid of $8049.0\pm2.0$\,\AA\,and assuming an intrinsic doublet centroid of $3728.5\pm0.5$\,\AA, this would imply that the GRB host lies at a redshift of $z=1.1588 \pm 0.0006$.  The width of the emission feature is consistent with an  [\ion{O}{2}] doublet at this redshift (3.19~\AA\ separation at $z=1.159$) when convolved with the instrumental resolution.  Furthermore, if the emission is from [\ion{O}{2}] we would not expect to see any other spectral lines; the most common, redder lines would fall redward of our red spectrum, and the only strong emission line blueward of [\ion{O}{2}] is Ly$\alpha$ which lies below our spectral coverage.  

Adopting the 8050~\AA\ emission feature as the [\ion{O}{2}] doublet, we can estimate the star-formation rate (SFR) of the host galaxy based on the line flux.  We measure the total flux from the extracted, one-dimensional spectrum in the region 8042.0--8056.0~\AA.  We measure the continuum by computing the median flux in two regions free of sky lines (regions ``C'' in Figure~\ref{fig:lrisspec}).  The total flux in these lines is $(1.6 \pm 0.2) \times 10^{-17}$ erg s$^{-1}$ cm$^{-2}$.  Using the relation between [\ion{O}{2}] luminosity and SFR described by \citet{Kennicutt1998}, we find that the GRB host galaxy is undergoing star formation at a rate of $1.6 \pm 0.2$ M$_{\odot}\ \rm yr^{-1}$ (assuming $H_{\circ}=71$ km s$^{-1}$ Mpc$^{-1}$, $\Omega_{\rm M}=0.3$, $\Omega_{\Lambda}=0.7$, as is done throughout this paper).  The intrinsic uncertainty in the calibration converting [\ion{O}{2}] line luminosity into SFR is approximately 30\% \citep{Kennicutt1998}; this uncertainty is not included in our quoted error.  Though our measured SFR must be considered a lower limit because we have not accounted for dust extinction, it is interesting to note that the rate of star formation in the host galaxy of GRB\,061126 is comparable to the SFRs found in other GRB hosts, measured using dust-corrected UV fluxes as well as [\ion{O}{2}] luminosities \citep{Christensen+2004}.

\section{Data Analysis and Modeling}

\subsection{Prompt Emission}
\label{sec:prompt}

\begin{figure*}[p] 
\centerline{\includegraphics[width=5.5in,angle=0]{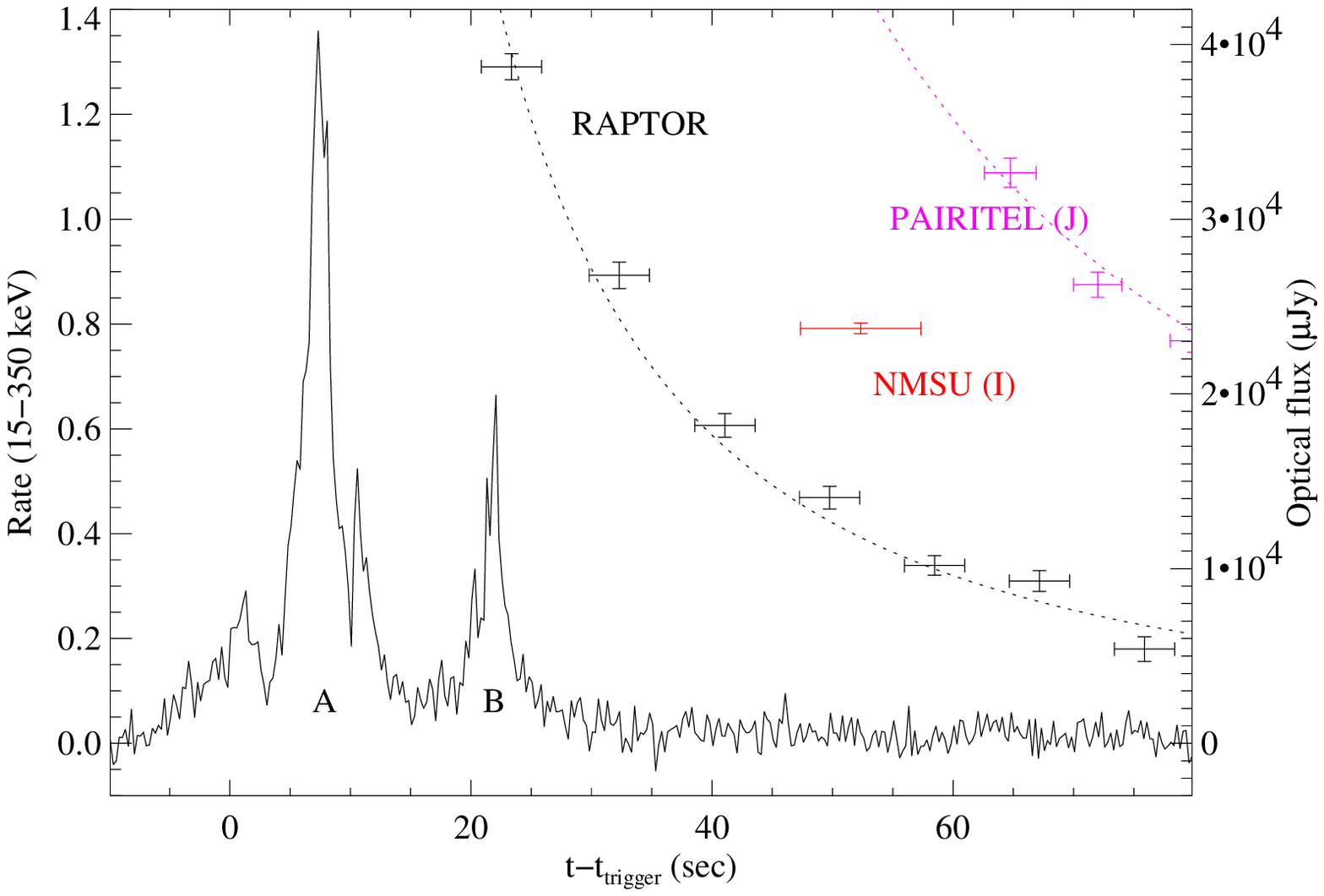}}
\caption[relation] {\small BAT 15--350 keV light curve of GRB\,061126, showing the profile of the main burst (dominated by two primary pulses, labeled ``A'' and ``B'') as well as the rapid response from the robotic telescopes RAPTOR,  the NMSU 1.0~m, and PAIRITEL.  At this early time the light curves are well fit by a simple power law  ($F_{\nu} \propto t^{-1.5}$) with $t_0$ simply set to the trigger time. There is no evidence  for a rising component or any correlation of the optical emission with the prompt emission.}
\label{fig:grblc}
\end{figure*}

The prompt emission from GRB\,061126 displays a complex, multi-peaked profile dominated by two large pulses (Figure~\ref{fig:grblc}).  After a gradually increasing component starting at 7~s before the trigger and a small precursor spike at $t \approx 1$~s after the trigger, the first and largest pulse (``Pulse A'') begins at $t \approx 3$~s, and fades to about twice the background level by $t \approx 15$~s.  A second pulse (``Pulse B'') begins at $t \approx 19$~s, lasting until $t \approx 25$~s.  There is short-timescale microstructure in both of these pulses. The full burst has a $T_{90}$ \citep{Kouveliotou+1993} of $26.8 \pm 0.8$~s in the full \emph{Swift} 15--350 keV band \citep{Butler+2007}.

\begin{figure*}[p] 
\centerline{\includegraphics[width=5.5in,angle=0]{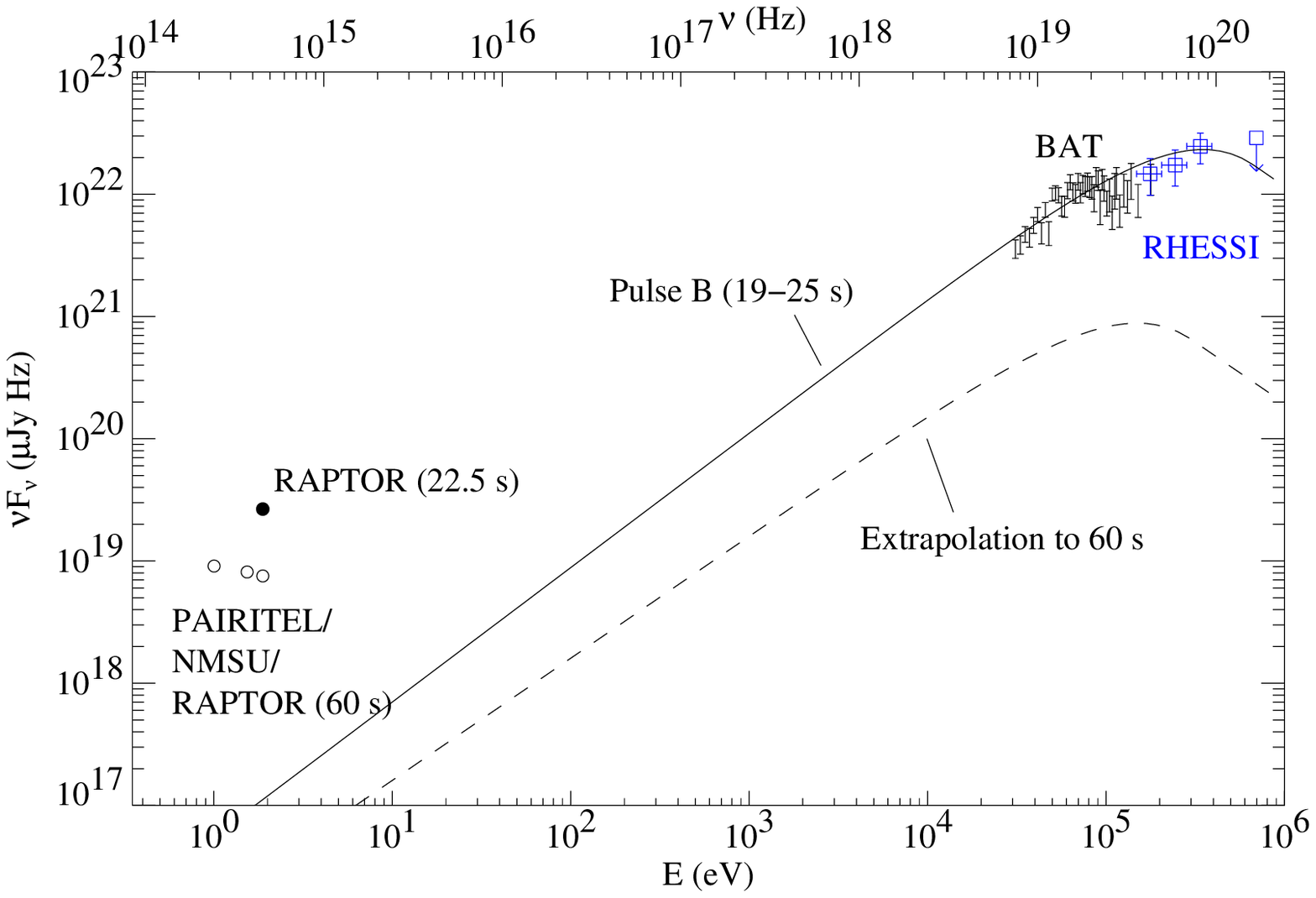}}
\caption[relation] {\small High-energy SED during the second main pulse of the prompt emission (``Pulse B''), with the contemporaneous RAPTOR measurement superimposed.  An extrapolation of our best-fit Band \citep{Band+1993} model (Table \ref{tab:spec_pars}) into the optical is shown to underpredict the optical flux by several orders of magnitude,  suggesting that even at this very early time the optical afterglow is already present at 20~s post-trigger, dominating the early-time flux at long wavelengths.  A temporal extrapolation of the BAT light curve to 60~s similarly underpredicts our multi-color data at that time.}
\label{fig:pulse2}
\end{figure*} 

Using the combined \emph{Swift} BAT and RHESSI spectrum, we fit a Band model \citep{Band+1993} over several different time regions:  The entire burst, the first pulse, and the second pulse.  Results are summarized in Table \ref{tab:spec_pars}.  For the full burst, we measure an average peak energy $E_{p,obs}$ of 620 keV, though there probably is hard-to-soft spectral evolution, as evidenced by the significantly different values of $E_{p,obs}$ during the two pulses.  The total fluence over the full spectral range is $(3.0\pm0.4) \times 10^{-5}$ erg cm$^{-2}$, which at the putative host redshift of 1.16 corresponds to an isotropic release of energy of $E_{iso} = (1.06\pm 0.14) \times 10^{53}$ erg over a 1--1000 keV host-frame bandpass, assuming our standard cosmology.

 Given this value of $E_{iso}$, the Amati relation \citep{Amati+2002} predicts $E_{p,obs} = 130 \times 10^{0.0\pm0.3}$ keV, which is 2$\sigma$ from the measured value.  Given $E_{iso}$ and $E_{p,obs}$, the Ghirlanda relation \citep{Ghirlanda+2004} predicts a jet break at $t=10^{1.41\pm0.27}$~d (13--49~d).  The beaming-corrected energy release is $10^{51.6\pm0.2}$ erg, for a jet opening angle of $10^{1.21\pm0.10}$ degrees (13--20$^\circ$).

RAPTOR detected GRB\,061126 contemporaneously with the BAT emission. Fortuitously, the first unfiltered exposure (which took place 20.87--25.87~s after the BAT trigger) matches quite well the second pulse of the GRB (peak time of 22.5~s and a $T_{90}$ of about 5~s) --- see the first optical point of Figure \ref{fig:grblc}.

Comparing this first, contemporaneous data point to later exposures in which the prompt emission has faded, the early-time RAPTOR data are seen to fade as a simple power law ($F_{\nu} \propto t^{-\alpha}$), with the decay index $\alpha \approx$ 1.5 (using the GRB trigger time as $t_0$). There is no evidence for an additional prompt flux component based on an extrapolation backward from later measurements.  Consistent with this observation, if we extrapolate the best-fit Band \citep{Band+1993} model of the second GRB pulse into the optical (Figure \ref{fig:pulse2}), the predicted optical flux is only 250~$\mu$Jy, significantly less than what is observed.  Finally, if we extrapolate the Band model of the prompt emission in time to 60~s (assuming continued  hard-to-soft spectral evolution and fading), it both falls far short of our multi-color data at that time and also has a different spectral slope  (open circles and dashed line in Figure \ref{fig:pulse2}).  For this burst, there is no clearly observable association between the prompt emission and the long-wavelength afterglow, even as early as 20~s after the burst.  This is not inconsistent with earlier reports of a link \citepeg{Blake+2005, Vestrand+2006, Yost+2007}: more likely, as our SED shows, for this burst the prompt component is simply dominated by an extremely bright early-time afterglow.

\subsection{Optical Light Curve}
\label{sec:optlc}

\begin{figure*}[p] 
\centerline{\includegraphics[width=5.5in,angle=0]{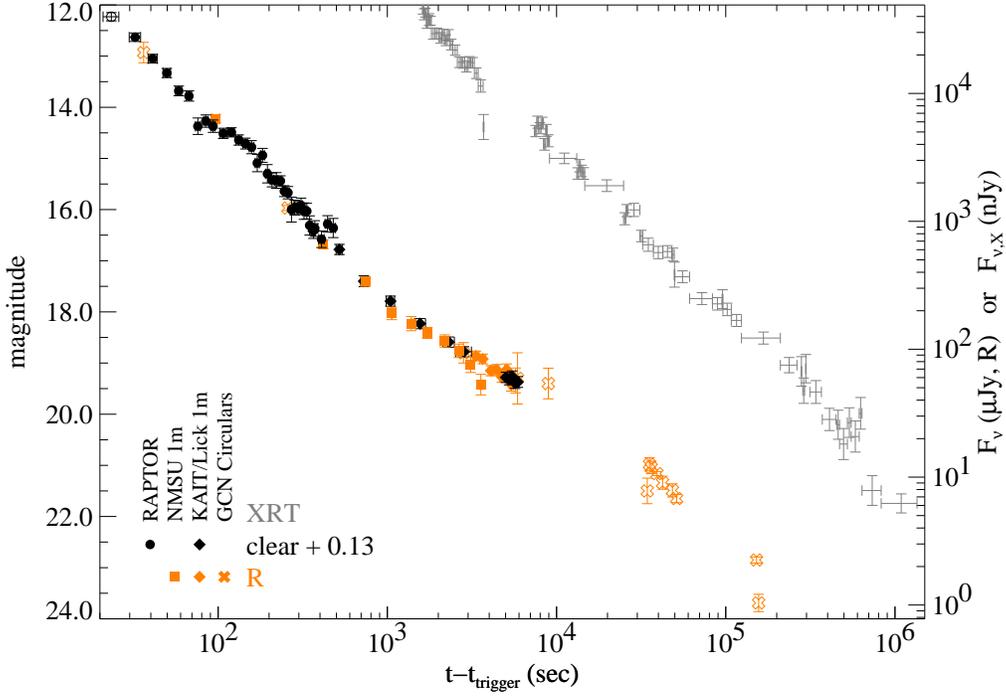}}
\caption[relation] {\small $R$-band and X-ray light curves of the afterglow of GRB\,061126 
showing the basic features of the early-to-late optical and X-ray afterglow light curves. 
Unfiltered data are also included, offset by 0.13 mag to match the $R$-band calibration.  Optical data are shown as filled (used in modeling) or unfilled (not used) symbols; X-ray data are shown as error bars with no central symbol.  There is a rapid decay with a bump at early times, transitioning to a significantly slower decay that probably breaks at late times.  Due to a delayed slew the X-ray afterglow was not observed until 1600~s, but from then until it faded below the detection threshold at $\sim9$~d it decays as an unbroken power law.  The 1 keV normalized flux is plotted.}
\label{fig:lcR}
\end{figure*}

\begin{figure*}[p] 
\centerline{\includegraphics[width=5.5in,angle=0]{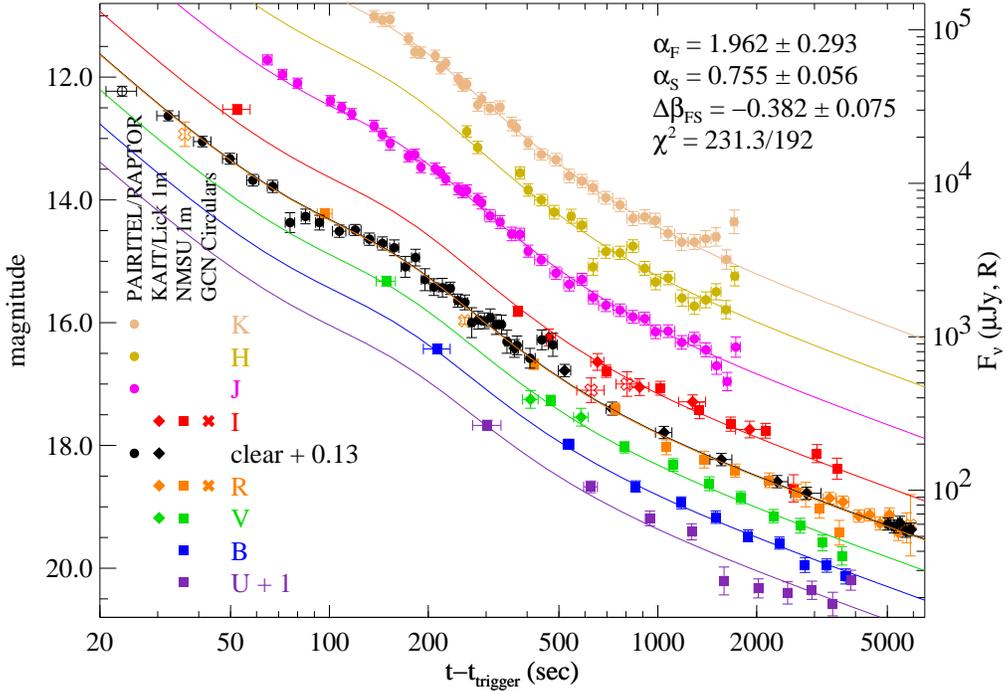}}
\caption[relation] {\small Broadband light curves of our early-time, multi-color photometry of the afterglow.  The light curves are fitted with a three-component broken power-law model, assuming that the third component (with the slowest decay) has a spectral index that differs from that of the early component by an amount $\Delta \beta$.}
\label{fig:lcmulti}
\end{figure*}

While the very early optical light curve appears to follow a power-law behavior, the light curve enters a more complex phase within a few minutes.  A brief visual inspection of the overall afterglow light curve (Figures \ref{fig:lcR} and \ref{fig:lcmulti}) shows several interesting features of this burst.   The early decay that began in our earliest data continues with the same trend for several decades in time, fading roughly as a power law with an average decay index $\alpha$ of about 1.5.  A simple power law is clearly a poor fit, however, due to the presence of of a prominent ``bump'' feature at around 120~s.

\begin{figure*}[p] 
\centerline{\includegraphics[width=6.5in,angle=0]{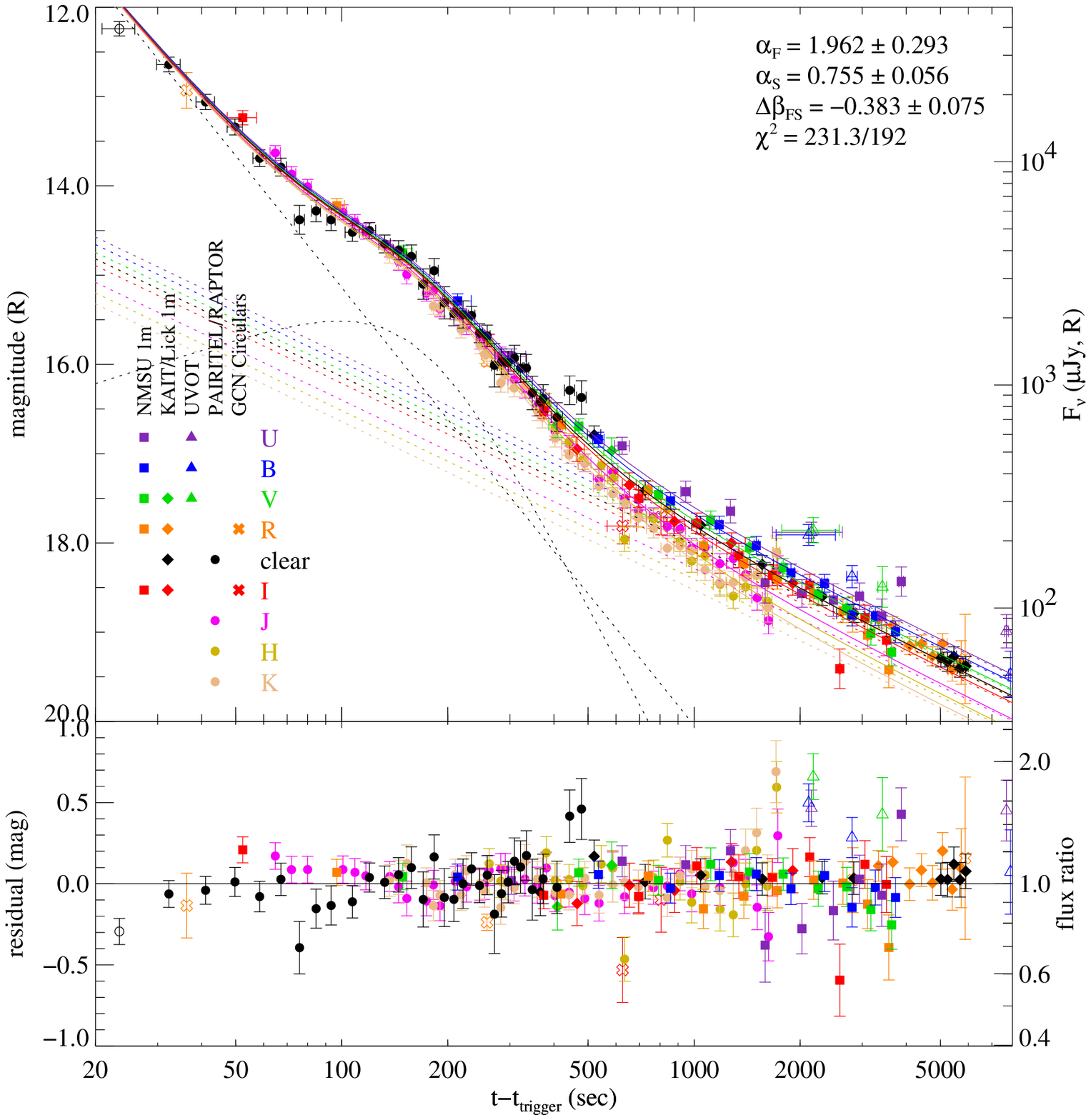}}
\caption[relation] {\small Broadband light curves of the afterglow with the curves aligned based on the early-time flux, emphasizing the red-to-blue color change.  The model is the same as in Figure \ref{fig:lcmulti}.}
\label{fig:lcppb}
\end{figure*}

Rapid early decay gives way to a shallower-decaying ($\alpha < 1$) component starting at about $10^3$~s that dominates for the rest of our observations.  Reports from the GCNs indicate an apparent steepening at later times, though with different authors \citep{GCN5876,GCN5902,GCN5903} disputing the value of the late-time decay index.

In our modeling, we construct the light curve as a sum of several broken power-law components, each of generally different color as well as different decay indices.  Details of the model follow.

The broken power-law model is that of \citet{Beuermann+1999}.  Mathematically, the fitted curve (as a function of the filter, represented by a central frequency $\nu$) has the form

$$F_{\nu}(t) = \sum_{j=0}^{n}  F_{\nu,j} \left[ \left( 
    \frac{t}{t_{\nu,j}} \right)^{{\alpha_{j,b} s}} +
   \left( 
    \frac{t}{t_{\nu,j}} \right)^{{\alpha_{j,a} s}}
   \right]^{-1/s} .
$$

\noindent
The flux offset $F_{\nu,j}$ generally depends on the filter (represented by the subscript $\nu$) as well as on the component $j$.  The temporal decay index $\alpha$ depends on the component and on whether it is before the break (subscript $b$) or after (subscript $a$), but does not depend on the frequency.  

In our modeling, we use three total summed components, though two of them are tightly linked:  a ``fast'' component, with no break ($\alpha_{F,a} = \alpha_{F,b}$), a ``bump'' component, fixed to have the same color as the fast component ($F_{\nu,B} \propto F_{\nu,F}$), and a ``slow'' component, also with no break ($\alpha_{S,a} = \alpha_{S,b}$).  

The constraints on the slow component depend on the specific model tested.  In all cases, the decay index ($\alpha_S$) is free to vary. We do not directly constrain the spectral index (defined here using the convention $F_{\nu} \propto \nu_{eff}^{-\beta}$) or require a power-law SED.  However, in some models we do constrain the change in the spectral properties.  In our different models, these color constraints are as follows:

\begin{itemize}
\item Unconstrained, allowing the linear flux factor $F_{\nu,S}$ to take arbitrary values for each filter.
\item Constrained to allow no spectral evolution: $F_{\nu,S} = C F_{\nu,F}$, where $C$ is a fitted constant.
\item Constrained in such a way as to permit the spectral index of the slow component to differ from that of the fast component by $\Delta \beta_{F-S}$ by fixing $F_{\nu,S} = C F_{\nu,F} (\nu/\nu_{R-band})^{-\Delta \beta_{F-S}}$.  Note that this {\it difference} in the spectral index can be fit directly regardless of Galactic extinction (which is significant) or host-galaxy extinction as long as the extinction can be assumed to be a constant in time.  
\end{itemize}

We also attempt two other, somewhat more specialized models:

\begin{itemize}
\item A fourth, physically motivated model in which the slow component is treated as an adiabatic forward shock whose flux peaks at some time during our multi-color observations.  In this case the slow component is broken, where the rising and decay indices ($\alpha_{S,b}$ and $\alpha_{S,a}$, respectively) and the change in spectral index across the break ($\Delta\beta_{S}$) are fixed using the constraints from the X-ray spectrum discussed in \S \ref{sec:closure}. 

\item  Solely for the purposes of estimating the ``average'' early-time optical decay index, we also fit a model with no bump component at all.  (This is a very poor fit, and because most of the IR observations were performed during the bump any photometric SED generated would be highly biased.  Therefore we do not use this model for any other purpose.)
\end{itemize}

Our data only trace the light curve with no gaps in coverage until $\sim 7000$~s, and we have no color information past 4000 seconds aside from a marginal $J$-band detection and $H$ and $K_s$ limits from the second night.  Later-time measurements are present in the GCN Circulars,  but different authors report different and somewhat contradictory behavior, suggesting either a problem with some of the public data or complex behavior of the light curve at late times. Consequently, we do not include any optical points after 10$^4$~s in our fits, anticipating that the late-time optical evolution will be discussed in greater detail in upcoming work by Mundell and collaborators.  Instead, we focus on the properties of the early and intermediate afterglow.
 
After fitting, we correct for Galactic extinction of $E_{B-V} = 0.182$ mag (using the NED extinction calculator\footnote{The NASA/IPAC Extragalactic Database (NED) is operated by the Jet Propulsion Laboratory, California Institute of Technology, under contract with the National Aeronautics and Space Administration.}, \citealt{Schlegel+1998}) and fit a simple power law to estimate the observed spectral index for different components.  (This neglects the possibility of host extinction, but as we will show in \S \ref{sec:bbfits}, if host extinction is present it does not cause significant deviation from a power law.)

\begin{figure*}[t] 
\centerline{\includegraphics[width=5.5in,angle=0]{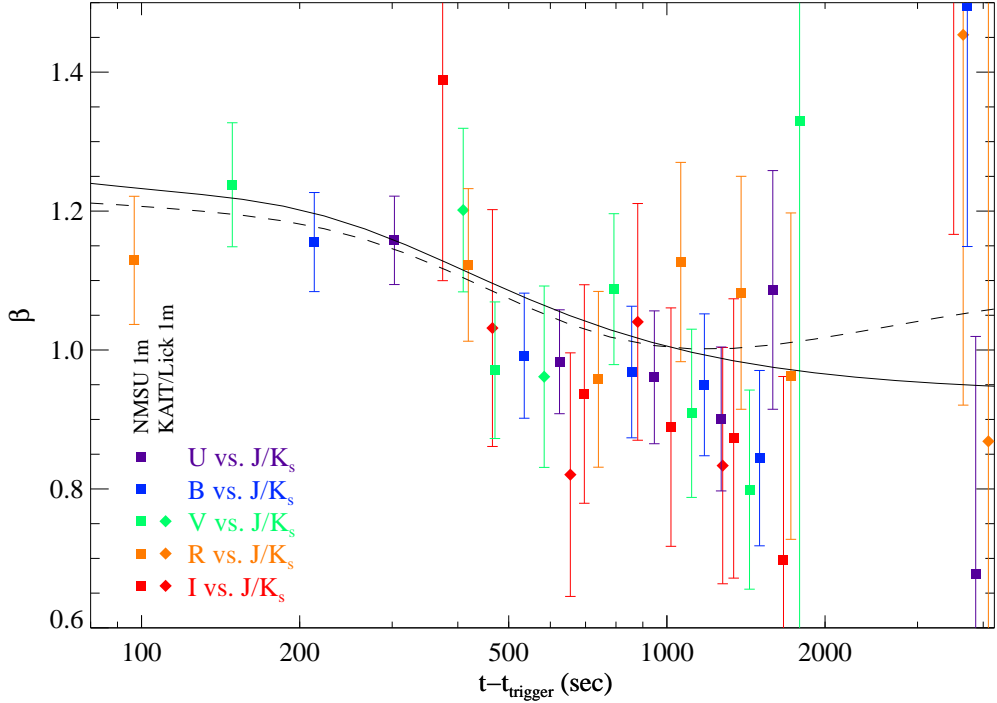}}
\caption[relation] {\small Evidence for color change across the transition from the rapidly decaying component to the slowly decaying component.  We fit a power law to simultaneous $J$/$K_s$ PAIRITEL exposures and optical exposures (in $U$, $B$, $V$, $R$, or $I$; color-coded appropriately) from the NMSU 1.0~m telescope.  The solid curves are not direct fits to these data, but represent the spectral index that would be observed at each time if fit to $K_s-U$ photometry based on two of our models.  The solid line is for a model where the late-time component of the afterglow is modeled as a simple power law; the dashed line represents a model of a forward shock undergoing a minimum-energy break at approximately
the transition time.}
\label{fig:betaplot}
\end{figure*} 

Despite the complexity of the models, no fit is observed to give a value of $\chi^2$ per degree of freedom (dof) $\approx 1$. The most successful model, using unbroken power laws and arbitrary filter-dependent color change, gives $\chi^2/$dof $\approx 6$.  This is not surprising; modulations in afterglow flux have been observed in many previous cases (for example, \citealt{Lipkin+2004}).  Accordingly, we add in quadrature an extra uncertainty of 0.08 mag to all the photometric measurements.  With this adjustment, our different models produce fit results as given in Table~\ref{tab:lcmodels}.  There is strong evidence for color change --- a fit allowing no change in the relative fluxes of different components has a value of $\chi^2/$dof = 261.9/193, while an equivalent fit allowing the early fast-fading and later slow-fading components to have different spectral indexes (Figure~\ref{fig:lcppb}) gives $\chi^2/$dof = 231.3/192.  (The $f$-test significance of this improvement is $10^{-6}$.) Under this model, $\beta$ shifts toward the blue across the transition by $\Delta\beta_{F-S} = -0.383 \pm 0.075$.

\subsection{Optical Color Evolution}
\label{sec:optcolor}

The above results indicate color evolution across the steep-to-shallow transition. To confirm this behavior and ensure this is not an artifact of the modeling, we construct truly contemporaneous SEDs using simultaneous PAIRITEL and NMSU 1.0~m observations by mosaicing only those individual PAIRITEL exposures taken during the NMSU exposure time range.  In cases where the exposure time is much less than the time since the burst, we increase the effective PAIRITEL exposure time by also including exposures shortly before the beginning or after the end of the optical exposures (otherwise the S/N for the late-time PAIRITEL measurements is quite low). After correcting for Galactic extinction, we fit the three $J$/$K_s$/optical data points with a basic power law, and determine the best-fit value and uncertainty for the spectral index $\beta$.  (We omit $H$-band measurements from this fit, since the variable transmission introduces a significantly larger scatter in that band compared to $J$ and $K_s$.) The results are plotted in Figure \ref{fig:betaplot}.

The early-time colors from the first NMSU filter cycle are consistent with a spectral index of $\beta = 1.2 \pm 0.1$, but starting at around 500~s the colors shift notably blueward and at later times the index is typically $\beta = 0.95 \pm 0.10$.  This provides model-independent support of our fit conclusion that the afterglow has undergone a color change.

\subsection{X-ray Light Curve and Spectrum}
\label{sec:xraylcspec}

The XRT light curve (Figure \ref{fig:lcR}) fades as a purely unbroken power law with a decay index $\alpha_X = 1.31 \pm 0.01$ over the entire span of the \emph{Swift} observations, from $\sim$2~ks out to nearly 10~d.  There is some suggestion that the decay rate is slightly faster during the first orbit, but the unbroken fit is still good ($\chi^2$/dof $= 668.2/550$) and joins with an unbroken extrapolation of the fading BAT light curve.

After removing the effect of neutral hydrogen\footnote{More accurately, the X-ray absorption is dominated by light metals, rather than hydrogen.  However, we will use the standard terminology and refer to the absorption column in terms of the hydrogen density.} absorption [both Galactic, for which we estimate $N_{\rm H,Galactic} = 0.103 \times 10^{22}$ cm$^{-2}$ from \citet{DickeyLockman1990}, and at the host redshift, for which we calculate a best-fit value of $N_{\rm H,host} = (1.1 \pm 0.3) \times 10^{22}$ cm$^{-2}$], the X-ray spectrum is a good fit ($\chi^2$/dof = 213.9/238) to a simple power law, with a spectral index $\beta_X = 1.00 \pm 0.07$.  There is no evidence for spectral evolution during the observations: using the X-ray hardness ratio (see \citealt{ButlerKocevski2007}); we constrain the change in the X-ray spectral slope to be $\Delta \beta_X <0.4$ (90\% confidence).   Our analysis is consistent with that from the most recent GCN report for this event \citep{GCNR16.2}. \footnote{Note, however, that the GCN report quotes the photon spectral index, instead of the flux spectral index, and that since they fit neutral hydrogen absorption only at $z=0$ instead of the host redshift, their $N_H$ excess is significantly less than our $N_{\rm H,host}$.}

\section{Discussion}
\label{sec:discussion}
\subsection {Constraints on Synchrotron Model Parameters}
\label{sec:closure}

Due to the complex behavior of the optical light curve and the possibility of host extinction, it is difficult to apply any constraints on the physical parameters of this burst ($p$, $\epsilon_B$, $\epsilon_E$, etc.) from the optical observations alone.   However, in principle the degeneracy can be broken using the XRT data.

The fits to our data unambiguously show that the optical flux and X-ray flux are decaying at different rates during the period when we have contemporaneous data in both bands (1800--6000~s): $\alpha_{opt} = 0.75 \pm 0.05$ and $\alpha_X = 1.31 \pm 0.01${\bf}.  Assuming a normal synchrotron spectrum \citep{Sari+1998}, this allows only two possibilities for the ordering of the spectral breaks: $\nu_c < \nu_{opt} < \nu_m < \nu_X $(fast cooling), or $\nu_m < \nu_{opt} < \nu_m < \nu_X$ (slow cooling).  In either case, the X-ray spectral index is predicted  to be $\beta = p/2$.  The observed spectral index is $\beta = 1.00 \pm 0.07$, so we can confidently calculate  an electron index of $p = 2.00 \pm 0.14$, the minimum value for a distribution that is unbroken at high energies.

However, this conclusion is problematic in light of the observed X-ray light curve.   Independent of environment, the decay index of emission from an expanding adiabatic  forward shock in the regime $\nu_c, \nu_m < \nu$ should obey $\alpha = 3\beta/2 - 1/2$, which for our spectral data would imply  $\alpha = 1.00 \pm 0.10$, disagreeing by about 3$\sigma$ from the measured value.  

The discrepancy is significant and indicates that at least one standard assumption does not hold.  We consider several possibilities: 

\begin{enumerate}

\item Our conclusion of $\nu_c, \nu_m < \nu_X$ is incorrect.  But, this is unlikely: no other ordering can produce optical and X-ray light curves that decay at different rates. Furthermore, the X-ray light curve remains unbroken out to very late times ($\gsim$ 10$^6$s).  Such a late cooling break in the X-ray band would imply an extremely low density ($\ll$ 1 cm$^{-3}$), inconsistent with an intragalactic environment and the high $N_H$ column density observed in the X-ray data.  (A column depth of 1 kpc predicts $n > 4$ cm$^{-3}$ if the gas is uniformly distributed throughout the galaxy.)

\item The afterglow evolution is dominated by radiative losses and has not yet transitioned to an adiabatic phase, even out to $\sim9$~d.  This case would predict that $\alpha = 12\beta/7 - 2/7 = 1.42 \pm 0.12$, within 1$\sigma$ of the observed value.  However, this requires that the synchrotron spectrum still be fast-cooling, which demands that the optical band be above the synchrotron critical frequency ($\nu_c < \nu_{opt} < \nu_m $) to produce a fading optical light curve in a constant-density medium.  The specific, predicted value of $\alpha = 4/7 = 0.57$ is not statistically consistent with our observations of the optical decay index in this region, but it is possible that the measurement may be confused somewhat by the presence of soft breaks or other systematic effects.

\item The X-ray light curve is affected by synchrotron self-Compton (SSC) losses, which would steepen the decay slightly relative to that predicted by a no-Compton model \citep{SariEsin2001}.  The amount of predicted steepening is, unfortunately, relatively low: $\Delta\alpha \approx 0.5 (p-2)/(4-p)$, which for the X-ray inferred value of $p = 2\beta = 2.0 \pm 0.14$ gives a constraint of the SSC-corrected decay index of $1.0 \pm 0.14$, still off by $2\sigma$ from the observed value of $1.31 \pm 0.01$.

\item The X-ray emission is not due to a forward shock, or not due to synchrotron radiation.
\end{enumerate}

Some combination of factors may also be at work.  In any event, the specific resolution to this problem is not critical at this stage, and we will proceed assuming $p=2$, which is independent of any assumptions about the burst environment and requires only that the source of the X-ray light be synchrotron radiation, where the X-ray band is above the cooling and minimum-energy breaks.

In addition to identifying the spectral regimes, we also seek to identify the burst environment (constant-density, wind-stratified, or a spreading jet). A spreading jet has a rapidly decaying light curve and can be immediately ruled out.  A wind can also be ruled out, since it would predict that the optical flux should decay faster than the X-ray flux, yet we observe the opposite. However, a constant-density (interstellar medium) environment is consistent with all of the qualitative features of the light curve.

\subsection{The Very Early Afterglow Decay --- A Reverse Shock?}
\label{sec:revshock}

For any assumption about the X-rays, the behavior of the fast-decaying optical component cannot be reproduced as originating from a forward shock.  The early optical decay index is $\alpha_F \approx 1.5$ if we fit a single component (no bump), though in our fits with a bump component the decay rate is steeper.  (However, if we interpret the bump as a density variation or other modulation of a single power law, rather than a truly separate component, the value of $\alpha_F \approx 1.5$ is most appropriate.  For our discussion of the interpretation, we use the constraint $\alpha \gsim$ 1.5, which is certainly true independent of the number of components used in our model.)

This decay is steeper than the $\alpha_X = 1.31$ decay slope observed in the X-rays at late times.  In no regime of the forward-shock interstellar-medium model can the longer-wavelength emission decay faster than the X-rays at the same or later times \citep{Sari+1998}.\footnote{It is possible in a wind-stratified medium, but even then, the observed value of $\alpha_F$ is too steep for $p=2$.}  Furthermore, the subsequent passage of $\nu_{m}$ or $\nu_{c}$ through the optical bands should produce breaks in the optical light curves which can only accelerate the temporal decay, not decelerate it.

The lack of flux excess in the first RAPTOR measurement, and the extrapolation of the Band model into the visible frequency range, argue against association with this emission to a prompt optical component from the internal shocks that (presumably) produced the high-energy emission.  Therefore, the most likely candidate to be responsible for this early optical emission is an external reverse shock which is produced when the forward shock begins to interact with the surrounding medium. Such an explanation has been previously used by several authors to explain observed peculiarities in the early-time light curves of a number of GRB afterglows \citep{SariPiran1999, KobayashiZhang2003, Wei2003, ShaoDai2005}. 

\citet{Kobayashi2000} derived the temporal behavior of reverse-shock light curves for two different regimes: the thick shell (corresponding to cases in which the reverse shock is able to become relativistic as it crosses back through the ejecta) and the thin shell (for cases where the shock always remains nonrelativistic).  The thick shell case is associated with relatively long bursts, and in particular predicts that the peak of the reverse-shock emission will happen at about the same time as the burst itself, whereas the thin-shell case predicts an optical peak after the GRB.  As the RAPTOR data indicate an optical light curve that is already rapidly declining during the prompt emission, we apply the thick-shell case here.  (However, the thin-shell case gives identical numerical predictions for the region of interest for $p=2$.)

Before the shock crossing time $T$, the reverse-shock light curve is expected to rise as $t^{1/2}$, which as already discussed is not observed.  The light curve peaks at $T$ and then decays:  for frequencies $\nu_m < \nu < \nu_c$ the flux decays as $t^{-(73p + 21)/96}$, while for $\nu < \nu_m$ it decays as $t^{17/36}$.   (For $\nu_c < \nu$ the light curve shuts off and rapidly dives to zero, and for fast cooling the light curve either declines as $\alpha = 17/36$ or is shut off after $T$.)

Observationally, the early-time average decay index is $\alpha_{F,av} \approx 1.52 \pm 0.02$ if we fit only a single component with no bump, though because of the bump feature the exact decay rate clearly varies; given this complexity, a rigorous statistical comparison with the simplistic predictions is not possible.  However, we can unambiguously rule out all cases except for the slow-cooling case in which $\nu_m < \nu < \nu_c$, where, if we use the value of $p$ from the late-time X-ray spectrum, we predict $\alpha = 1.74 \pm 0.11$, which is generally consistent with our observations.

Our early-time multi-color photometry provides us with a rare opportunity to test the spectral predictions of reverse-shock models, in addition to the time-domain predictions.  The synchrotron spectrum of a reverse shock is predicted to be identical in its general form to that from a forward shock, with the exception that after the reverse shock crosses the shell (at time $T$), there is no emission from frequencies above some cutoff frequency $\nu_{\rm cut}$.  The spectral power-law indices are the same (for the same value of $p$), although the break frequencies $\nu_m$ and $\nu_c$ are generally not.

For the reverse shock, the only regime consistent with the data is $\nu_m < \nu < \nu_c$.  The prediction for the spectral index in this region (from \citealt{Sari+1998}) is $\nu^{-(p-1)/2}$.  Again assuming $p = 2.00 \pm 0.14$ from the late-time X-ray observations, we predict that the spectral index should be $\beta = 0.50 \pm 0.07$.  This is significantly less than the observed value (after correcting for Galactic extinction) of $\beta_F \approx 1.25$.  This very large discrepancy may be due to host-frame extinction.  However, as we will discuss in \S \ref{sec:bbfits}, the lack of observable curvature in the SEDs of either the fast-decaying or the slow-decaying components impose strong constraints on the amount and nature of any extinction.

\subsection{Transition to Forward Shock}
\label{sec:optlate}

The fast-decaying early-time optical component is subsumed by the shallower late-time component at around $10^3$~s.  The natural inclination is to assume that the reverse-shock emission has been replaced by that from a forward shock.

As already discussed, the different decay rates of the X-ray and optical light curves rule out the possibility that they are in the same synchrotron regime, so we expect the optical band to be below a spectral break.  For adiabatic evolution, if the optical band is below the cooling break, we expect (assuming $p=2.0$) an optical decay index of $\alpha_{opt} = 0.75$.  This is consistent with the data, both in the empirical model where we assume a power law that fades throughout the observations (for which we calculate $\alpha_{S} = 0.75 \pm 0.06$) and the more realistic fit in which the slow component rises, experiences a (chromatic) break, and then fades ($\chi^2$/dof $= 228.0/197$ for this model, slightly better than the best model assuming an unbroken slow decay.)  For radiative evolution and the case that the optical band is below the minimum-energy break (but above the cooling break and the critical frequency), $\alpha_{opt} = 4/7$, which is only consistent within 2.5$\sigma$.  

If the slower decay component is due to an adiabatic forward shock and the optical band is in the range $\nu_m < \nu_{opt} < \nu_c$ then the predicted intrinsic optical spectrum is $F_{\nu} \propto \nu^{-(p-1)/2}$, which for $p=2.0 \pm 0.14$ suggests $\beta = 0.5 \pm 0.07$.  If the forward shock is radiative, the prediction is the same but with no statistical uncertainty ($\beta$ = 1/2).  The observed spectral index of the possible forward-shock component is somewhat redder than this value, $\beta_S$ = $0.92 \pm 0.08$ --- or, using our physical model, the post-peak $\beta_S$ = $1.11 \pm 0.09$.  Again, one may appeal to host-galaxy extinction to make up the difference.

We will examine this possibility in greater detail in the next section.  For now, however, we can sidestep the host extinction question completely if we look only at the \emph{change} in spectral index between the early (reverse shock) and late (forward shock) components, as long as we can assume that the extinction is constant with time.   

As discussed earlier, the only reverse-shock model consistent with the observed early-time light curve requires that either $\nu_m < \nu_{opt} < \nu_c$ (adiabatic) or $\nu_c < \nu_{opt} < \nu_m$ (radiative), both of which predict the same value for the spectral index for $p \approx 2$.  We reach the same conclusion with the later-time data, meaning that the reverse and forward shocks are in the same synchrotron regime and should therefore have the same color as long as $p$ is the same.

However, we do in fact observe a color change between the two components, of $\Delta \beta = -0.38 \pm 0.08$ in our model.  There are several possible interpretations for this.

\begin{figure*}[p] 
\centerline{\includegraphics[width=5.5in,angle=0]{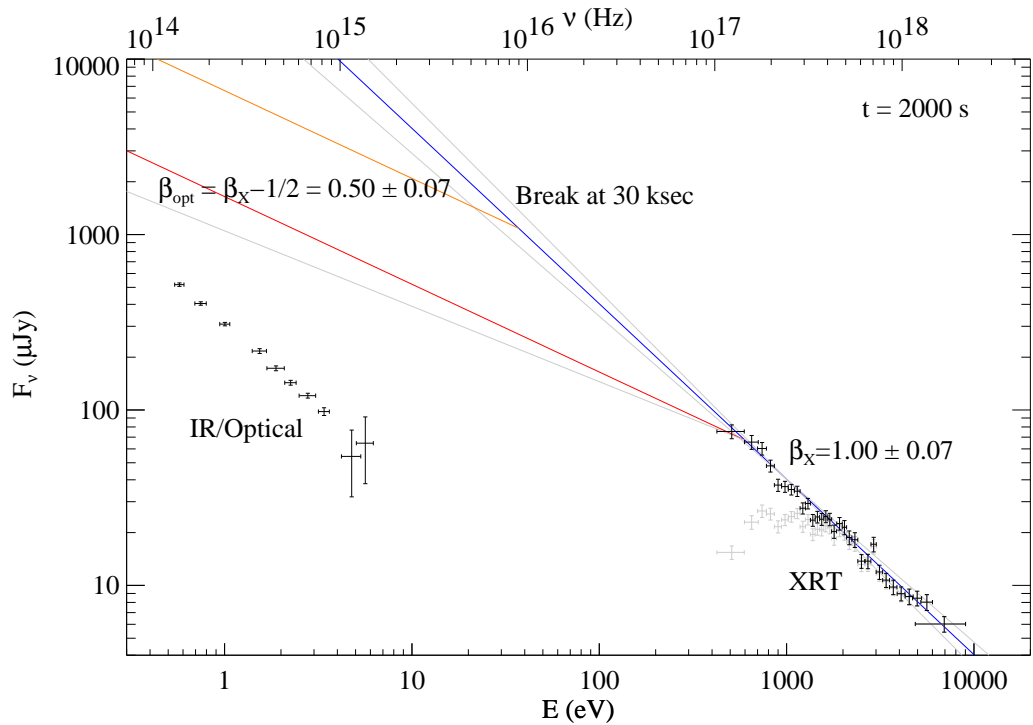}}
\caption[relation] {\small Broadband SED from optical to X-ray at $t=2000$~s after
the trigger.  The steep X-ray decay and apparent shallow optical decay places the cooling break between the X-ray and optical bands.  Even if we make the maximally generous assumption and place the cooling break at 1 keV, the optical flux is seen to be overpredicted (red line) by a factor of about $5$.   If we are less generous with our assumption, and choose to interpret the late-time break in the optical afterglow seen in the GCN Circulars as the effect of the spectral break passing through the $R$ band at that time, the discrepancy is even larger.  The optical data are a good fit to a power law, and it is difficult to appeal to extinction to make up the difference.  The black X-ray points are corrected for Galactic absorption plus a best-fit model of the host hydrogen column; grey points are corrected for Galactic absorption only (no host-galaxy absorption).  Optical data are corrected for Galactic extinction only.}
\label{fig:sedxo}
\end{figure*} 

Most likely, the forward shock is not strictly in the $\nu_m < \nu_{opt} < \nu_c$ regime at the transition time, but is in fact still in the process of breaking from its very blue rising portion (that is, $\nu_m \approx \nu_{opt} < \nu_c$), and does not attain its ``true'' color until significantly later, when it will shift back to the reverse-shock color.  Our data are quite consistent with this possibility; the physically motivated fit which models the reverse shock as a rising and falling with the appropriate change in spectral index over its peak is equally sound as the fit assuming no rising portion; the ultimate change in spectral index in this model is $\Delta\beta_{F-S,b} = -0.12 \pm 0.08$, consistent with no change within 2$\sigma$.  If this is the case, we predict that the afterglow color will shift back towards the red after the end of our observations.

Alternatively, one of our assumptions could be in error.  It is possible that the value of $p$ is different between the forward and reverse shocks; in this case a difference of  $p_{rev} - p_{fwd} = 0.7 \pm 0.1$ would be required.  Also, the amount of extinction could in fact be variable due to dust destruction by the GRB, though the general constancy of the early-time spectral index in Figure \ref{fig:betaplot} before the transition time seems to speak against this possibility.  Finally, if the optical or X-ray emission is strongly affected by another process such as Compton scattering, this could also produce spectral variability.

Of course, a combination of two or more of these factors is also possible.

\subsection{Broadband Spectral Fits and Constraints on Extinction}
\label{sec:bbfits}

The light curve fits performed in our analysis naturally give rise to spectral fits.  The filter-dependent flux parameters $F_{\nu,j}$ give the relative fluxes in each filter for both components of the light curve, and can be used to calculate the observed SED and look for signs of host extinction.  We have already referred to the best-fit spectral indices $\beta$ for the fast and slow components, all of which were fit assuming no host extinction.  Here we will use different extinction models to constrain the properties of any host-frame dust in greater detail.

Large amounts of extinction are implied by the X-ray to optical SED.  In Figure~\ref{fig:sedxo} we plot the SED of the ``slow'' component as computed at $t = 2000$~s, shortly before the end of our multi-filter observations and after the beginning of the XRT observations.\footnote{
Our physical model of the light curve in terms of a reverse-to-forward shock transition indicates that the intrinsic SED of the slow component may be variable due to the passage of a shallow minimum-energy break somewhat before this time, and so changes slightly after the end of our optical observations. However, to keep the discussion on firm observational footing without extrapolating our measurements beyond the range of our available data or favoring any particular model, we choose the simple empirical model (with no rising component) for discussion of extinction.  These results are not significantly affected by choosing the physical model instead.}
A model for Galactic extinction of $E(B-V) = 0.182$ mag has been subtracted.  The predicted X-ray to optical slope, and the predicted slope in the optical--IR frequency window, is $\beta=(p-1)/2 = 0.5$ in the adiabatic case and also $\beta=1/2 = 0.5$ in the radiative case, so the prediction is the same.  In fact, however, we measure a nearly flat X-ray to optical slope of $\beta_{OX} = 0.23$ (using the $R$-band and 1 keV fluxes), and an IR-optical slope of $\beta_{opt} \approx 0.95$.  This value of $\beta_{OX}$ is enough to unambiguously label this event as a ``dark burst'' by the criterion of \citet{Jakobsson+2004} (that is, any burst with $\beta_{OX}$ less than the $p=2$ synchrotron limit of 0.5) at this time, in spite of this being in fact one of the optically brightest bursts ever observed by \emph{Swift}!  This surprising fact is a consequence of a combination of the unusual late-time X-ray brightness and the rapid early fading of the optical afterglow.  (We could, perhaps, incorporate both this burst's early-time brightness and its later-time faintness with the moniker ``grey burst,'' since ``dark burst'' seems inappropriate to the burst's entire evolution.)

\begin{figure*}[p] 
\centerline{\includegraphics[width=5.5in,angle=0]{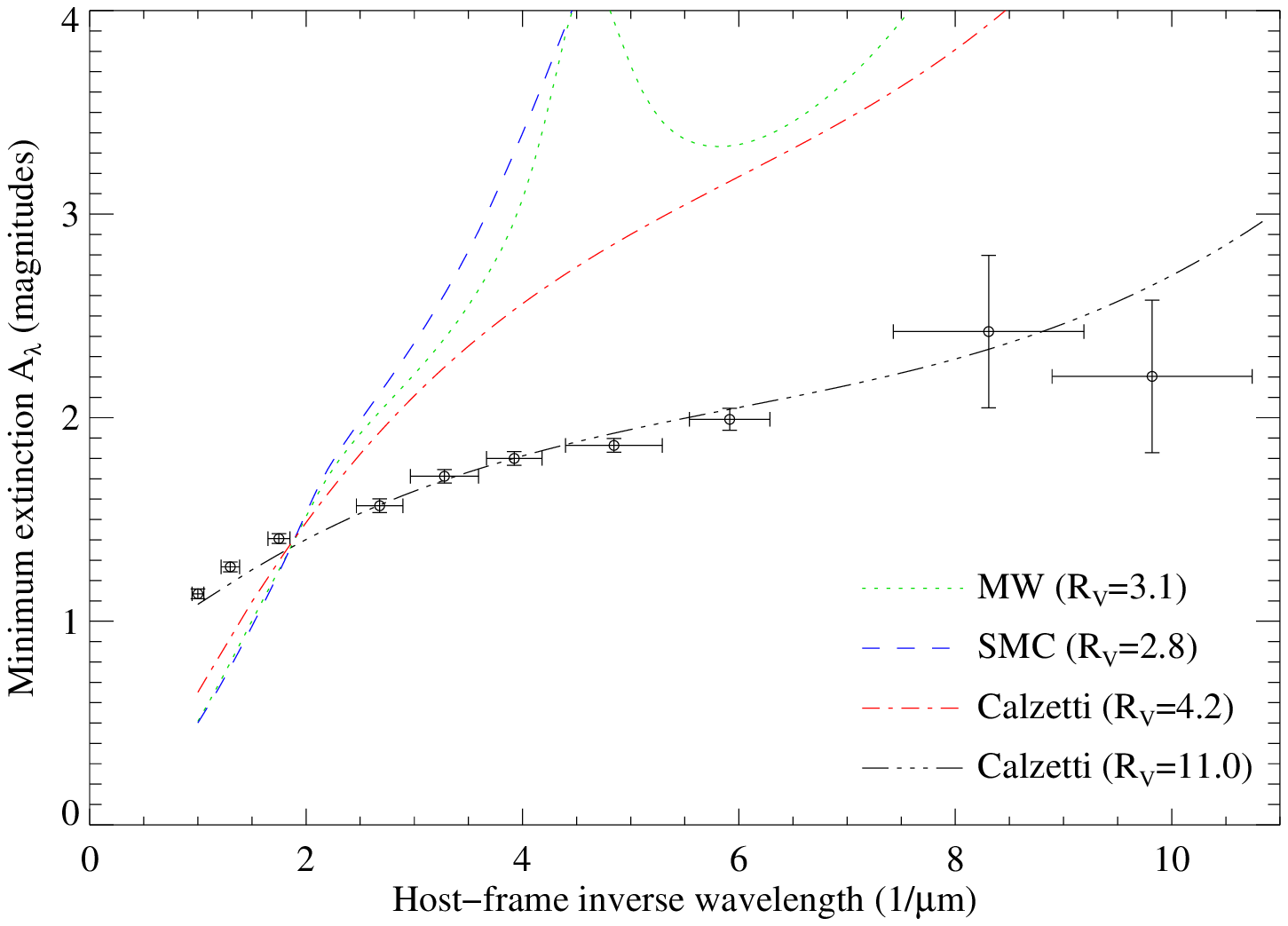}}
\caption[relation] {\small Plot of the minimum amount of extinction necessary in each band to resolve the discrepancy with the X-ray flux measurements (assuming $p=2$), compared to several representative extinction curves.  $A_V$ is fixed at 1.35 mag for all models.  The Milky Way and SMC models use the formulation of \citet{FitzpatrickMassa1990}; the Calzetti model is a model for extinction in starbursting galaxies from \citet{Calzetti+2000}.}
\label{fig:extcurves}
\end{figure*} 

The most commonly invoked interpretation for these dark bursts, and for X-ray/optical flux mismatches generally (e.g., \citealt{Schady+2007}), is that the optical flux has been suppressed by dust extinction (or, for very high-redshift bursts, hydrogen absorption).  A large amount of extinction would be necessary to apply this explanation here.  To calculate a minimum amount of extinction, we assume the case that the cooling break (or for radiative evolution, the minimum-energy break) is just redward of the X-ray band, and take the minimum value of the intrinsic $\beta_{OX} = 0.43$ permitted by our uncertainty in $\beta_X$ (this would imply $p <$ 2, which is within our errors).  For this case, the optical flux is overpredicted by more than a factor of 5.3 in the observed $V$ band, requiring $> 1.8$ mag of host extinction in this band, or $A_V=1$ mag in the host-galaxy frame.  This is an absolute minimum: requiring an intrinsic $\beta_{OX}=0.5$ ($p$ = 2 or a radiative regime) increases this to 2.2 mag in the observed $V$, and placing the break at a more general point between the X-ray and optical bands 
increases it even further.

Assuming a Milky Way metallicity and extinction law, our measurement of the host-frame hydrogen column of $N_{\rm H,host} = (1.1 \pm 0.3) \times 10^{22}$ cm$^{-2}$ would correspond to a large host-frame extinction column of $A_V \approx 6$ mag, even greater than that indicated by the X-ray/optical discrepancy.  However, this assumption is unlikely to be appropriate.  For a more realistic choice of the gas-to-dust ratio in the Small Magellanic Cloud (SMC; about 8.4 times the Galactic value; \citealt{Gordon+2003}) or a starburst galaxy ($\sim$ 10 times Galactic; \citealt{Lisenfeld+2002}), we expect a proportional reduction in $A_V$, predicting $A_V \approx 0.6$--0.9 mag, roughly consistent with the minimum necessary suppression.

The wide frequency coverage, photometric accuracy, and contemporaneous nature of our optical data gives us the opportunity to firmly constrain the extinction properties of  this event.  As such, we fit a power law plus an extinction component to our optical--IR SED, allowing the intrinsic unabsorbed spectral index $\beta$ to vary as a free parameter. We try numerous extinction profiles, including the Fitzpatrick \& Massa (``FM") model \citep{FitzpatrickMassa1990}, which has parameterizations for a wide range of galaxy types, of which we fit both a Milky Way extinction profile and a SMC profile using the parameters measured by \citet{Gordon+2003}.  We also fit the Pei extinction profile \citep{Pei1992} for the SMC, and the Calzetti extinction profile \citep{Calzetti+2000} measured from observations of starburst galaxies.  We use the flux parameters from our model of the slow component, and assume a systematic error of $2\%$ in the IR filters, $3\%$ in the optical filters, and $5\%$ in U-band.  The different extinction curves are plotted in Figure \ref{fig:extcurves}.

\begin{figure*}[p] 
\centerline{\includegraphics[width=5.5in,angle=0]{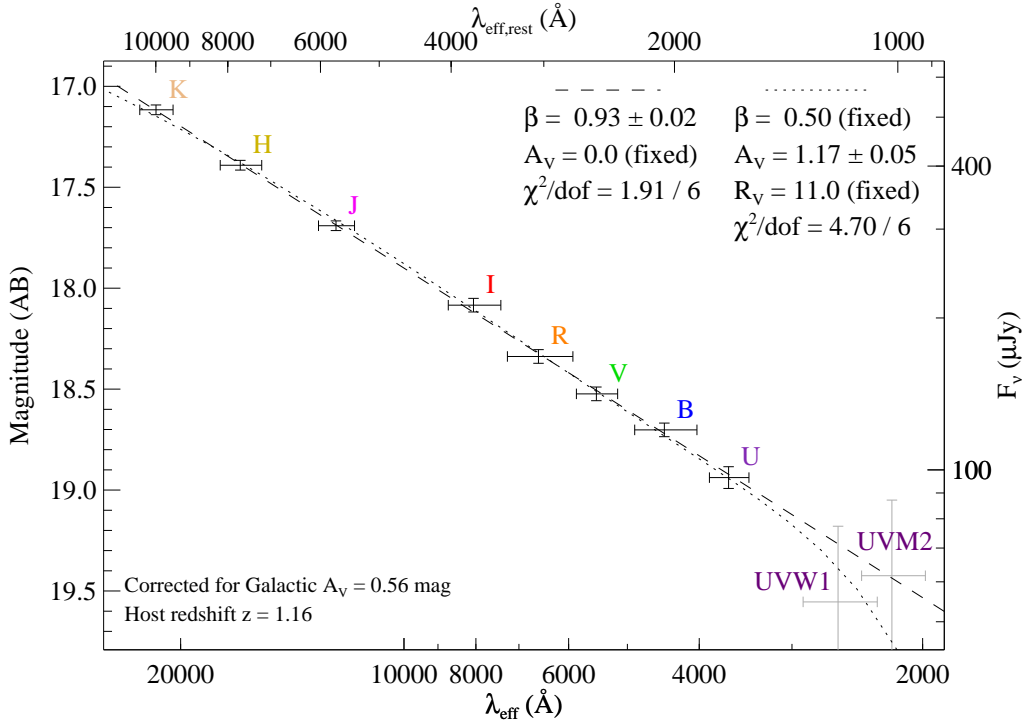}}
\caption[relation] {\small Spectral energy distribution of the slowly decaying component of the afterglow at 2000~s, fit both to a model assuming no host-galaxy extinction (dashed line) and a model assuming host-frame extinction is present in sufficient quantity to provide the observed minimum discrepancy between these optical measurements and the X-rays.  We use the Calzetti et al. (2000) extinction model of starburst galaxies, with $R_V$ in this case set very high, to 11.  The uncertainties of the UVOT measurements are very large and we do not include them in our fits.}
\label{fig:iroptsed}
\end{figure*}

In every case, the best-fit model is that of no host-frame extinction ($A_V$ = 0 mag; our fits restrict $A_V$ to be positive) or a very low value consistent with 0, indicating an intrinsic early-time spectral index equal to the observed spectral index of $\beta=0.93\pm0.02$. This extinction-free fit is good, with $\chi^2/$dof = $1.91/6$, and is plotted in Figure \ref{fig:iroptsed}.

\begin{figure*}[t] 
\centerline{\includegraphics[width=4.0in,angle=0]{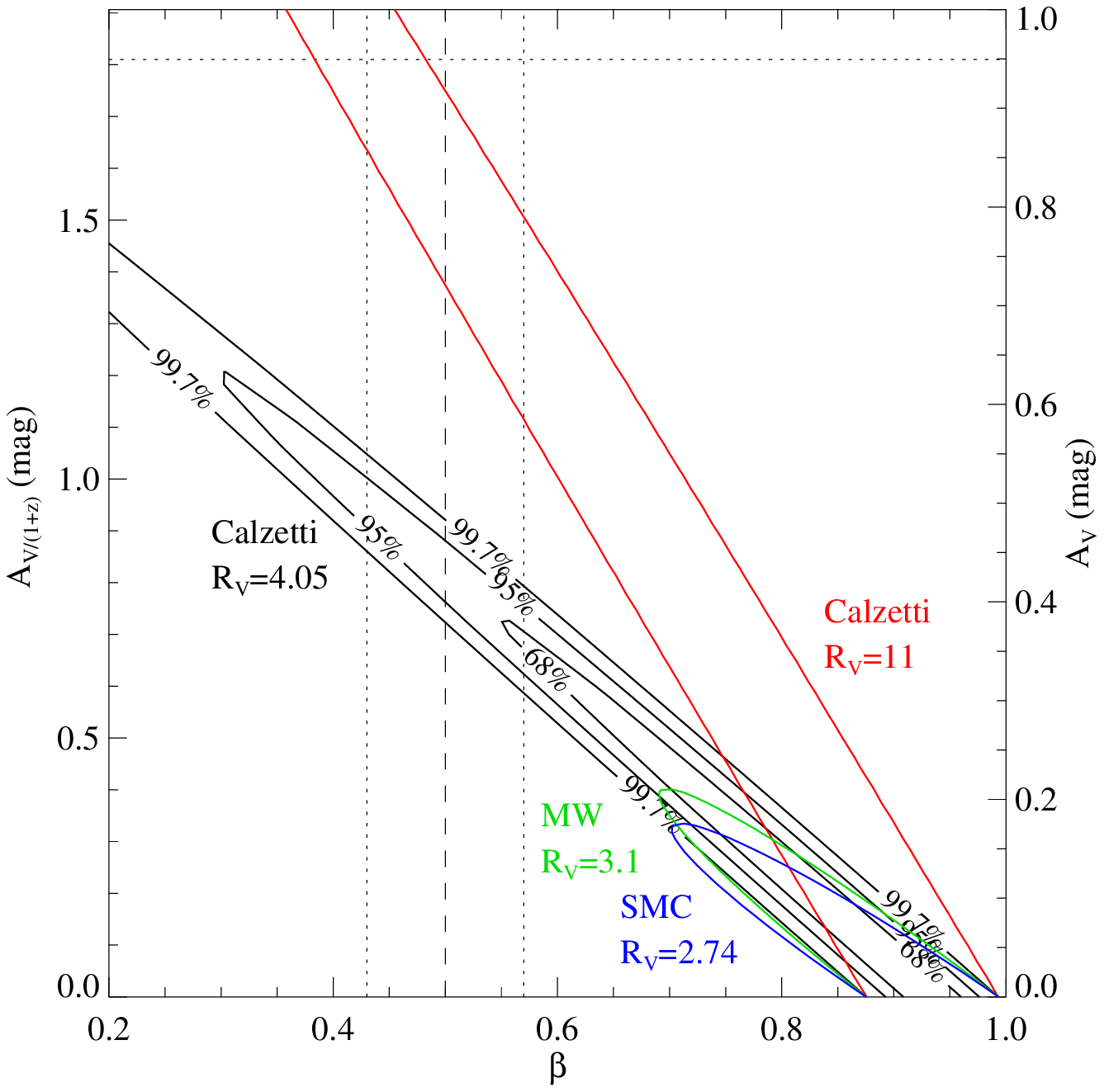}}
\caption[relation] {\small Plot of $\chi^2$ as a function of $A_V$ and the intrinsic
unabsorbed $\beta$ for different models.  Extinction is strongly limited for standard extinction laws.  The Calzetti et al. (2000) extinction law allows slightly larger amounts of extinction, but still short of the amount required (an absolute minimum of 1.8 mag, represented by the dotted line near the top).  The vertical bars indicate the range of $\beta$ predicted by the X-ray observations if the optical emission is due to a forward shock, though if the forward shock is still peaking the intrinsic $\beta$ is allowed to be blueward (lower $\beta$) of this range.}
\label{fig:extchisq}
\end{figure*}

The limit on $A_V$ depends largely on the model adopted: greyer dust-extinction models (and larger values of $R_V$, indicating larger grain sizes, within those models) permit more $A_V$.  However, all ``normal'' dust-model fits strictly limit the observed extinction to $A < 0.2$ mag in the observed $V$ band.  The relatively grey Calzetti model allows more extinction, but even for this case, the amount of observed extinction can only be achieved for extremely large ratios of the total-to-selective extinction ($R_V \gsim\ 8$) and the fit is degraded, $\chi^2/$dof $= 4.70/6$ (Figure \ref{fig:iroptsed}), though still acceptable.  The $\chi^2$ confidence contours for a few different fit models are plotted in Figure~\ref{fig:extchisq}.

This is, furthermore, for the contrived ``optimistic'' case where the X-rays are just above the break; for the more general case where the break is at lower energy, the required extinction is greatly increased.  There is evidence to believe that this break is relatively near the optical at this time --- the observations in the GCN Circulars suggest a break around $t \approx 3\times10^4$ s, and despite some discrepancies all observations (including our single-epoch late-time PAIRITEL observations) agree that the optical flux  by the second night has fallen well below the predicted value from the $\alpha < 1$ decay we measure at intermediate times.  If we assume that this break in the light curve were due to a spectral break passing through the $R$ band, at $t = 2000$~s the break would be at $\sim$40 eV, predicting an optical flux 4 times (1.5 mag) higher even than the above ``minimum'' prediction would suggest, or 3.3 mag total ($A_V \approx 1.8$ in the host frame.)

\section{Implications and Alternative Models}
\label{sec:impl} 

The standard assumptions of the fireball synchrotron model fare poorly when confronted with the available early-time data on GRB\,061126.  The most striking failure is the large discrepancy between optical and X-ray fluxes, though the unexplained bump feature in the early optical light curve and the lack of agreement between theoretical predictions for $\alpha$ and $\beta$ in X-rays are also unsettling.

It is beyond the scope of this paper to fully resolve this quandary.  Nevertheless, we present some possible solutions and briefly analyze their relative merits and failings in light of our available observations.

\subsection{Grey Dust?}
\label{sec:greydust}

We have demonstrated that no standard extinction law can fully explain the discrepancy between the X-ray and optical fluxes at late times.  Sufficiently high values of $R_V$ (8--12) --- corresponding to relatively grey dust laws, and physically to large grain sizes --- do provide a respectable fit, but since even in the starburst galaxies studied by Calzetti, and in the molecular clouds in the Milky Way, $R_V$ is generally in the range of only 2--5, it is worth asking if this is physically reasonable.

There is ample reason to suspect that GRB host galaxies, and the GRB progenitor sites within those galaxies, will have dust properties that do not resemble the local universe.  GRBs in general (and GRB\,061126 is no exception) occur at high redshift and in very low-mass, high star-formation galaxies, and dust properties may evolve with time and almost certainly do correlate with host-galaxy type.  Furthermore, if GRB progenitors are very massive stars, they should occur near their birth sites (probably within dense clouds); in the Milky Way, such environments are known to be correlated with anomalous extinction curves and high values of $R_V$ \citep{Cardelli+1988}.  Finally, the GRB itself is expected to destroy dust within several parsecs of its progenitor, which can preferentially affect certain types of dust --- for example, destroy smaller grains while leaving the larger grains relatively untouched \citep{WaxmanDraine2000, Perna+2003}.

Furthermore, this is not the first GRB in which the observed optical flux has been deficient relative to the X-ray flux, or relative to a late-time synchrotron prediction using the optical decay rate.  In many cases a relatively normal extinction law [except, in nearly all cases, for the lack of the characteristic 2175~\AA\,bump feature seen in the Milky Way and the Large Magellanic Cloud (LMC)] has successfully explained the suppression \citepeg{Schady+2007}, although we caution that without broadband photometry (including IR measurements) it can be difficult to accurately distinguish different models, and extinction laws other than the standard Milky Way, LMC, and SMC laws are often not tested.  Even so, in many other cases \citepeg{Stratta+2005,Chen+2006} the low-redshift dust models clearly do not work --- in these cases, the optical data are consistent with an unabsorbed power-law, but one that disagrees with an extrapolation from the X-ray band.  Most of these authors favor a grey-dust interpretation of one manner or another.  Unusual reddening laws have also been reported from observations of supernovae \citep{Wang+2006, Elias-Rosa+2006} and high-redshift quasars \citep{Maiolino+2004}.

Unusual dust probably remains the simplest explanation for the flux discrepancy.  Nevertheless, at least two factors give us pause when considering this explanation.
First, our fits to the optical data alone simply do not require it.  There is no physically compelling reason that we can think of why dust should exhibit an extinction law that simply transforms one power law to another, requiring $A_{\lambda} \propto (0.5 \pm 0.1) \log(\nu/\nu_o) + C$ or something mimicking it within our observational uncertainties.  From the optical data alone, it is more natural to assume that the excellent power-law SED is intrinsic, and look elsewhere to explain the disagreement with the X-ray data.

Second, the observed dust-to-gas ratio, while seeming to favor large amounts of extinction, actually greatly overpredicts the extinction if we assume a physical interpretation of the relatively grey extinction law in terms of dust that is mostly bound up in large grains.  The grain size distributions presented by \citet{WeingartnerDraine2001}, for example, are already biased (as weighted by the grain volume) toward the largest grains, so for the same dust-to-gas ratio, skewing the grain distribution further toward large grains will significantly decrease the opacity at long wavelengths but also decrease it at short wavelengths.  When the anticipated low dust-to-gas ratio of the GRB host galaxy is considered, the observed X-ray $N_H$ column is already only just enough to explain the observed extinction; suppressing it further by binding the available dust into very large grains becomes problematic. 

Nevertheless, without a detailed physical modeling of the potential dust in the GRB environment (including the possible effects of time-variable opacity due to dust destruction), we can neither firmly confirm nor rule out the presence of grey dust in this burst.  Such an examination is beyond the scope of our present analysis; however, we encourage dust modelers to make use of our observations of the minimum extinction required  (Figure \ref{fig:extcurves}) to help determine the viability of grey-dust models for this event.  Our early-time infrared and optical photometry should strongly constrain any models involving dust destruction.

\subsection{Synchrotron Self-Compton as the Origin of the X-ray Afterglow}
\label{sec:compton}

We now turn to interpretations in which the optical SED is treated as intrinsic, and instead look to processes primarily affecting the X-rays to explain the anomalous behavior.  The discrepancy between the X-ray $\alpha$ and $\beta$ independently argues that the X-ray emission for this burst may not obey the normal assumptions about GRB afterglows.

The SED in Figure \ref{fig:sedxo}, if intrinsic, has two peaks, reminiscent of the spectrum created by inverse-Compton scattering from a synchrotron source.  Furthermore, it is interesting that the observed optical spectral index is consistent with the X-ray spectral index:  perhaps the X-ray afterglow is dominated by inverse synchrotron self-Compton flux boosted from the optical band.  Compton scattering has not commonly been invoked in interpreting GRB afterglow observations, but it has been used to explain X-ray/optical flux discrepancies similar to this one in at least two previously published cases: GRB\,000926 \citep{Harrison+2001} and GRB\,030227 \citep{CastroTirado+2003}.

However, we consider SSC unlikely to be the solution to our discrepancy for several reasons.

\begin{enumerate}
\item Even admitting for the effect of Compton scattering on the light curve, the optical data do not obey the predictions of any model without an extinction correction.  The optical and X-ray have very similar spectral slopes ($\beta_{opt} = 0.93 \pm 0.02$,  $\beta_X = 1.00 \pm 0.07$), suggesting that the X-ray photons are in fact upscattered optical photons, and that (nongray) extinction is minimal, so this value of $\beta$ must be treated as intrinsic.  This restricts us to two synchrotron regimes:  $\nu_{c} < \nu_{opt}$ or  $\nu_{m} < \nu_{opt} < \nu_{c}$.  The former case would imply $p=2$ and requires a light curve that falls off as $\alpha = (3p-2)/4 + (p-2)/(2(4-p)) = 1.0$, which is ruled out by the observations.  (The optical light curve decays quite slowly: $\alpha = 0.75 \pm 0.06$ during our observations, though it steepens later.)  The latter case would imply $p=3$, and predict $\alpha = 3(p-1)/4 = 1.5$, which is also ruled out.

\item Inverse-Compton scattering does not produce a broken power-law SED at high energies; \citet{SariEsin2001} have shown that in reality the breaks are significantly softened, in disagreement with the X-ray observations which indicate a simple power-law spectral slope.

\item Finally, it is difficult to construct the Compton-scattered portion of the broadband SED without ``contaminating'' the optical bands with Compton-scattered flux from lower energies in a way that would noticeably bias the blue end of our photometric SED.  (However, this effect may be alleviated somewhat if the Compton-boosted self-absorption frequency $\nu_{a,IC}$ is not much less than the Compton-boosted cooling frequency $\nu_{c,IC}$, since this would make the low-energy side of the Compton-scattered spectrum much steeper than what is otherwise a very slowly falling ($\beta = -1/3$) low-energy tail.)
\end{enumerate}

\subsection{Physically Separate X-Ray and Optical Emission Regions}
\label{sec:separate}

Finally, we consider the possibility that the emission regions for the optical and X-ray emission are spatially distinct.  There may be two forward shocks (each peaking in a different energy range) resulting from a two-component jet \citep{Peng+2005}, or the entire optical afterglow may be due to a reverse shock with an erratic profile extending out to late times (as in \citealt{UhmBeloborodov2007}), while the X-ray light curve may be due to a more well-behaved radiative forward shock.

The notion of finding a way to physically separate the optical and X-ray emission has recently found support in other contexts, in particular to try to resolve the apparent lack of correlation between \emph{Swift} X-ray and optical breaks \citep{Oates+2007} and to explain \emph{Swift} X-ray breaks generally \citep{Panaitescu+2006}.  Unfortunately, the relatively short period of overlap between the optical and X-ray data for GRB\,061126 prevents us from being able to constrain this possibility in any detail for this event.  However, we will note that any two-origin explanation shares the difficulty just discussed in the context of whether inverse Compton may explain the X-ray flux excess --- that is, how to avoid contaminating the blue end of the observed optical spectrum with emission from the low-energy tail of a synchrotron-like spectrum peaking at X-ray wavelengths.

\section{Conclusions}
\label{sec:conc}

We have presented multi-wavelength, early-time observations of the recent, bright GRB\,061126.
In our favored model, the early optical data appear to be well explained by a reverse shock.  The reverse-shock emission fades rapidly and is dominated by emission from the forward shock from $\sim$500~s onward.  A small color change is observed at this transition, likely due to the peak of the forward-shock synchrotron spectrum passing through the optical band at roughly the same time.

However, only with significant contrivance can this model also explain the X-ray observations from the \emph{Swift} XRT, which are much brighter than an extrapolation of the optical flux would predict.  The optical flux may be suppressed by host-galaxy extinction, but this would require a remarkably grey dust law, capable of generating 1.5--3 mag of optical extinction without causing the observed $U$--$K_s$ SED to deviate from a power law.  Alternatively, the X-ray and optical emission may be due to two physically or spatially distinct emission processes, but our observations strongly restrict the positions and sharpness of any spectral breaks and cast doubt on the viability of this model.

In either case, we advise caution in interpreting future early afterglows in the absence of data sets that are not well sampled both spectrally and temporally.  Many previous studies, by necessity, have been restricted in either their temporal properties (sampling), frequency coverage (available filters and wavebands), or both. Under the standard fireball model these assumptions seemed valid, as only a single parameter set would fit them.  Here we show that when a more complete picture is available, no parameter set seems to fit the data, unless we invoke large quantities of grey dust or separate origins for the X-ray and optical afterglows.

The very bright X-ray afterglow and late-time optical faintness are enough to qualify this event under some definitions as a dark gamma-ray burst in spite of the extraordinary bright early afterglow.  Had the event been intrinsically fainter or at higher redshift, or had the optical follow-up observations been delayed significantly, the optical afterglow may have been missed entirely.  Therefore, it is possible that events like GRB\,061126 may represent a significant fraction of dark bursts.

The most common interpretation of dark bursts is that they are due to absorption of the optical flux by host-galaxy dust or by neutral hydrogen at very high redshift.   Our study presents tentative evidence for a third possibility, which is that the optical  faintness may be intrinsic to the GRB itself, due to enhancement of the X-ray flux or intrinsic suppression of optical flux, or both.   If the optical faintness is due to dust, our results suggest that near-IR and optical observations alone may not be as constraining as once hoped, since any extinction must not cause significant deviations from a power law, even across a decade in frequency from the near-IR to the UV.  However, the combination of near-IR, optical, and X-ray observations remains a potent tool for understanding GRBs, and we anticipate that additional multi-wavelength observations in the coming years will continue to shed light on ``anomalous'' events like GRB\,061126.

\acknowledgements

The RHESSI team is supported by NASA/\emph{Swift} AO-2 Guest Investigator grant NNG06GH58G.  We thank Kevin Hurley and Claudia Wigger for their support in acquiring and analyzing these data. 
RAPTOR is primarily supported by Laboratory Directed Research and Development funding at Los Alamos National Laboratory.
PAIRITEL is operated by the Smithsonian Astrophysical Observatory (SAO) and was made possible by a grant from the Harvard University Milton Fund, a camera loan from the University of Virginia, and continued support of the SAO and UC Berkeley. The PAIRITEL project and D.P. are further supported by NASA/\emph{Swift} Guest Investigator grant NNG06GH50G. We thank M.\ Skrutskie for his continued support of the PAIRITEL project.  
KAIT was made possible by donations from Sun Microsystems, Inc., the Hewlett-Packard Company, AutoScope Corporation, Lick Observatory, the National Science Foundation, the University of California, the Sylvia \& Jim Katzman Foundation, and the TABASGO Foundation. A.V.F.'s group at U.C. Berkeley is supported by NSF grant AST--0607485 and the TABASGO Foundation, as well as by NASA/\emph{Swift} Guest Investigator grant NNG05GF35G.
Some of the data presented herein were obtained at the W.\ M.\ Keck Observatory, which is operated as a scientific partnership among the California Institute of Technology, the University of California, and NASA; the Observatory was made possible by the generous financial support of the W.\ M.\ Keck Foundation. We wish to extend special thanks to those of Hawaiian ancestry on whose sacred mountain we are privileged to be guests.  
We are grateful to Erin McMahon for her critique of the synchrotron model presented in our first version of this manuscript, and to the anonymous referee for a very thorough and helpful commentary.

\bibliographystyle{apj}

\clearpage

\begin{deluxetable}{llclll}
\tabletypesize{\small}
\tablecaption{GCN Observations of GRB\,061126\label{tab:gcnphot}}
\tablecolumns{6}
\tablehead{
\colhead{GCN} &
\colhead{$\Delta t$\tablenotemark{a}} & \colhead{Filter} &
\colhead{Magnitude\tablenotemark{b}} & \colhead{Flux\tablenotemark{b}} & \colhead{Reference} \\
\colhead{} & \colhead{(s)} & \colhead{} &
\colhead{} &
\colhead{($\mu$Jy)} & \colhead{}}
\startdata
5869 &   36.3 & $R$ & $ 12.93 \pm  0.20$ & $20893 \pm  3515$ & (1)\\
5857 &  258.3 & $R$ & $ 15.97 \pm  0.05$ & $ 1270.6 \pm    57.2$ & (2)\\
5868 &  626.4 & $I$ & $ 17.10 \pm  0.20$ & $ 351.6 \pm  59.1$ & (3)\\
5868 &  806.1 & $I$ & $ 17.00 \pm  0.20$ & $ 385.5 \pm  64.9$ & (3)\\
5859 & 2820.1 & $R$ & $ 18.80 \pm  0.20$ & $   93.8 \pm    15.8$ & (4)\\
5859 & 5880.4 & $R$ & $ 19.30 \pm  0.50$ & $   59.2 \pm    21.8$ & (4)\\
5859 & 8939.8 & $R$ & $ 19.40 \pm  0.30$ & $   54.0 \pm    13.0$ & (4)\\
5866 &34367.3 & $R$ & $ 21.50 \pm  0.25$ & $    7.8 \pm     1.6$ & (5)\\
5876 &152064 & $R$ & $ 22.85 \pm  0.06$  & $    2.2 \pm     0.1$ & (6)\\
5875 &156381 & $R$ & $ 23.69 \pm  0.17$  & $    1.0 \pm     0.2$ & (7)\\
5902 &39225.6 & $R$ & $ 21.16 \pm  0.04$ & $   10.7 \pm     0.4$ & (8)\\
5902 &51235.2 & $R$ & $ 21.65 \pm  0.08$ & $    6.8 \pm     0.5$ & (8)\\
5903 &35424.0 & $R$ & $ 20.98 \pm  0.10$ & $   12.6 \pm     1.1$ & (9)\\
5903 &36288.0 & $R$ & $ 21.04 \pm  0.09$ & $   11.9 \pm     0.9$ & (9)\\
5903 &42336.0 & $R$ & $ 21.34 \pm  0.10$ & $    9.0 \pm     0.8$ & (9)\\
5903 &48384.0 & $R$ & $ 21.49 \pm  0.10$ & $    7.9 \pm     0.7$ & (9)\\
\enddata 
\tablecomments{GCN data points were not used in our light-curve models.}
\tablenotetext{a}{Exposure mid-time, measured from the \emph{Swift} trigger (UT 08:47:56.4).}
\tablenotetext{b}{Observed value; not corrected for Galactic extinction.}
\tablecomments{References. -- (1) \citealt{GCN5869};
             (2) \citealt{GCN5857};
             (3) \citealt{GCN5868};
             (4) \citealt{GCN5859};
             (5) \citealt{GCN5866};
             (6) \citealt{GCN5876};
             (7) \citealt{GCN5875};
             (8) \citealt{GCN5902};
             (9) \citealt{GCN5903}}
\end{deluxetable}

\begin{deluxetable}{llllll}
\tabletypesize{\small}
\tablewidth{0pt}
\tablecaption{Results of Band et al. (1993) model fits to the BAT$+$RHESSI spectrum of GRB~061126}
\tablehead{ \colhead{Region} & \colhead{$\alpha$} & \colhead{$\beta$} & \colhead{$E_{\rm peak}^{\rm obs}$} & \colhead{100 keV Norm.} & \colhead{$\chi^{2}_{\nu}$ (DOF)} \\
\colhead{} & \colhead{} & \colhead{} & \colhead{[keV]} & \colhead{[ph cm$^{-2}$ s$^{-1}$] } }
\startdata
Full ($t=-6$--$35$s) & $-1.06\pm 0.07$ & $<-2.3$ & $620^{+220}_{-160}$ & $7.8^{+0.7}_{-0.5} \times 10^{-3}$ & 0.733 (106)\\

Pulse A ($t=3$--$14$s) & $-0.94\pm 0.06$ & $<-2.5$ & $790^{+160}_{-130}$ & $1.8\pm 0.1 \times 10^{-2}$ & 1.042 (105)\\

Pulse B ($t=19$--$25$s) & $-0.9^{+0.2}_{-0.1}$  & ... & $350^{+190}_{-110}$ & $1.1^{+0.4}_{-0.1} \times 10^{-2}$ & 1.179 (100)\\

\enddata
\tablecomments{The quoted uncertainties correspond to the 90\% confidence region.  The data in each time region are acceptably fit by an exponential times a power-law model.  The high-energy power-law component (with photon index $\beta$) is not required in the fits but can be constrained for regions ``Full'' and ``Pulse A.'' Using $\alpha\approx -1$ and the declining $E_{\rm peak}$ and normalization values between pulses A and B, we estimate an approximately energy-independent GRB spectral flux of $0.1^{+0.1}_{-0.5}$ mJy below 1 keV at $t=60$~s (Figure \ref{fig:pulse2}).
}
\label{tab:spec_pars}
\end{deluxetable}

\begin{deluxetable}{llllll}
\tablewidth{0pc}
\tablecaption{Optical Light-Curve Fits}
\tablehead{ \colhead{$\alpha_F$} & \colhead{$\beta_F$} & \colhead{$\alpha_{S,b}$} & \colhead{$\alpha_{S,a}$} & \colhead{$\Delta\beta_{F-S}$} & \colhead{$\chi^2$/dof}}
\startdata
$2.09\pm0.29$ & $1.28\pm0.01$ & -    & $0.80\pm0.05$ & $-0.32\pm0.03\tablenotemark{a}$  &  212.0/185 \\
$1.76\pm0.22$ & $1.07\pm0.02$ & -    & $0.58\pm0.12$ & $ 0\tablenotemark{b}$    &  261.8/193 \\
$1.96\pm0.29$ & $1.31\pm0.02$ & -    & $0.75\pm0.06$ & $-0.38\pm0.08$  &  231.3/192 \\  
$1.70\pm0.09$ & $1.23\pm0.02$ & $-$0.50\tablenotemark{b}& $0.75\tablenotemark{b}$ & $-0.12\pm0.09$\tablenotemark{c}
                                                                       &  227.6/197  \\ 
\enddata
\tablecomments{Summary of key parameters and $\chi^2$ from the various models fit to the data.  The model with no color change is strongly ruled out. The nature of the color change depends on the assumed model.}
\tablenotetext{a}{Not a formal fit parameter in this model --- the flux amplitude parameters in each filter are allowed to assume their arbitrary best-fit values.  In other models the change in these parameters is constrained to be due to variation in the spectral index $\beta$.}
\tablenotetext{b}{Fixed parameter.}
\tablenotetext{c}{Change between the spectral index of the fast component and the index fit to the slow component after its peak ($\beta_{S,b}$).  The slow component undergoes a chromatic break from $\beta_{S,b}$ = 0.284 to $\beta_{S,b}$ = 1.11.}
\label{tab:lcmodels}
\end{deluxetable}

\end{document}